\documentclass[prd,floats,superscriptaddress,nofootinbib,preprintnumbers,twocolumn]{revtex4}%
\usepackage{lipsum, babel}

\usepackage{amssymb}
\usepackage{amsmath}

\usepackage{sans}

\makeatletter\AtBeginDocument{\let\@elt\relax}\makeatother

\usepackage{graphicx}
\usepackage{graphics}
\usepackage{dcolumn}
\usepackage{color}
\usepackage{rotate}
\usepackage{fancyhdr} 
\usepackage{hyperref}
\usepackage{indentfirst}
\usepackage{booktabs} % LT for tables
\usepackage{fontawesome} % LT for github logo
\usepackage{physics} % LT for useful routines
\usepackage{siunitx} % For SI stuff
\DeclareSIUnit \parsec {pc}
\DeclareSIUnit \year {yr}
\usepackage{umoline}

\def\be{\begin{eqnarray}}
\def\ee{\end{eqnarray}}
\def\ba{\begin{eqnarray}}
\def\ea{\end{eqnarray}}
\def\no{\nonumber}

\definecolor{darkred}{rgb}{.743,0,0}

\begin{document}
\title{Gravitational lensing $H_0$ tension from ultralight axion galactic cores}

\author{Kfir Blum}\email{kfir.blum@weizmann.ac.il}
\affiliation{Weizmann Institute of Science, Rehovot 7610001, Israel} 
 \author{Luca Teodori}\email{luca.teodori@weizmann.ac.il}  
 \affiliation{Weizmann Institute of Science, Rehovot 7610001, Israel} 
%\date{\today}

\begin{abstract}
Gravitational lensing time delays offer an avenue to measure the Hubble parameter $H_0$, with some analyses suggesting a tension with early-type probes of $H_0$. 
The lensing measurements must mitigate systematic uncertainties due to the mass modelling of lens galaxies. In particular, a core component in the lens density profile would form an approximate local mass sheet degeneracy and could bias $H_0$ in the right direction to solve the lensing tension. We consider ultralight dark matter as a possible mechanism to generate such galactic cores. We show that cores of roughly the required properties could arise naturally if an ultralight axion of mass $m\sim10^{-25}$~eV makes up a fraction of order ten percent of the total cosmological dark matter density. A relic abundance of this order of magnitude could come from vacuum misalignment. Stellar kinematics measurements of well-resolved massive galaxies (including the Milky Way) may offer a way to test the scenario. Kinematics analyses aiming to test the core hypothesis in massive elliptical lens galaxies should not, in general, adopt the perfect mass sheet limit, as ignoring the finite extent of an actual physical core could lead to significant systematic errors.  
\end{abstract}
%\pacs{}

\maketitle

%\tableofcontents
%%%%%%%%%%%%%%%%%%%
\section{Introduction}
Measurements of the image and time delay of gravitationally-lensed quasar-host galaxies constrain the expansion rate of the Universe, parameterised via the Hubble constant $H_0$~\cite{Refsdal64,Kochanek_2002,Kochanek06}. 
In a work that summarised the efforts of several groups, the TDCOSMO team\footnote{\url{http://www.tdcosmo.org/}} used these data to derive $H_0=74.0^{+1.7}_{-1.8}$~km/s/Mpc (TDCOSMO-I~\cite{Millon:2019slk}). 
This result is in tension with measurements based on the cosmic microwave background (CMB)~\cite{Akrami:2018vks}, which find $H_0 = 67.36 \pm 0.54$ km/s/Mpc, and with large scale structure (LSS) galaxy clustering that is consistent with the CMB~\cite{DESH0,Ivanov:2019pdj,DAmico:2019fhj,Troster:2019ean}. We refer to the apparent discrepancy between the lensing~\cite{Millon:2019slk} and the CMB/LSS~\cite{Akrami:2018vks,DESH0,Ivanov:2019pdj,DAmico:2019fhj,Troster:2019ean} measurements as the {\it lensing $H_0$ tension}. 

The lensing $H_0$ measurement of Ref.~\cite{Millon:2019slk} is independent of the well-known cepheid-calibrated supernova-Ia (SNIa) measurements by the SH0ES collaboration, which find $H_0=73.2\pm1.3$~km/s/Mpc~\cite{Riess:2020fzl}. The lensing result~\cite{Millon:2019slk} is in excellent agreement with the SNIa/cepheids result~\cite{Riess:2020fzl} and both are ``late Universe" probes of $H_0$, that is, they involve only low-redshift ($z\sim1$) dynamics, in contrast to the CMB/LSS measurements which can be considered ``early Universe" probes because they hinge crucially on high-redshift ($z\sim10^3$) dynamics such as the baryonic perturbations sound horizon. 
Discrepancy between early and late determinations of $H_0$ could indicate a long-awaited breakdown of the $\Lambda$CDM effective description of cosmology~\cite{Verde:2019ivm,DiValentino:2021izs}. After all, we understand no more than 5\% of the energy budget of the Universe. It is tantalising to think that a clue to the nature of the remaining 95\% may come from the $H_0$ tension.

Needless to say, all of the methods to determine $H_0$ require a careful account of systematic uncertainties. A main concern in the SNIa analyses is the calibration of local distance ladder anchors. The TRGB-calibrated SNIa analysis of Ref.~\cite{Freedman:2020dne}, for example, finds a value of $H_0$ that is consistent to $\sim1\sigma$ with the CMB result, despite a nominal precision that is comparable to the SNIa/cepheids method. (See, however,~\cite{Soltis:2020gpl}. And of course, there are concerns of systematic issues in the CMB analysis, too~\cite{DiValentino:2021izs}.) 

\begin{figure}[h!]
\centering
%\hspace*{-0.8cm}
 \includegraphics[scale=0.125]{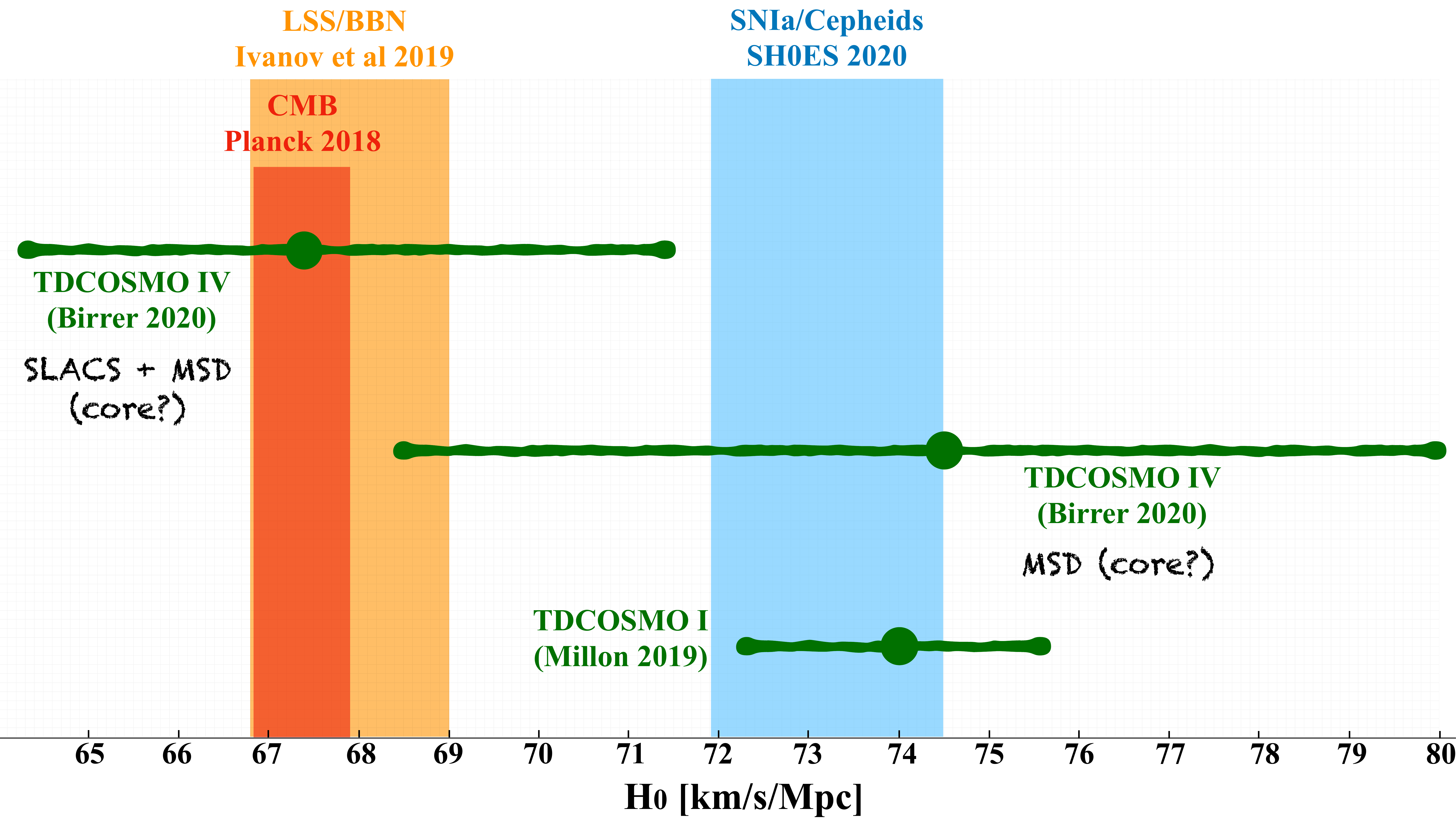}
 \caption{Status of the lensing $H_0$ measurements. The ``internal MSD" degree of freedom, added in moving from TDCOSMO-I~\cite{Millon:2019slk} to TDCOSMO-IV~\cite{Birrer:2020tax}, relaxed the lensing constraint on $H_0$. But what is the physical explanation of the internal MSD component? 
 }
 \label{fig:H0}
\end{figure}

Lensing measurements of $H_0$ are detached from the distance ladder. However, modelling degeneracies couple the inferred value of $H_0$ to the assumed density profile of the lens galaxy~\cite{Falco1985,Schneider:2013sxa,Schneider:2013wga,Xu:2015dra,Birrer:2015fsm,Unruh:2016adf,Tagore:2017pir,Sonnenfeld:2017dca,GomerWilliams19,Kochanek:2019ruu,Kochanek:2002rk}. 
Ref.~\cite{Blum:2020mgu} pointed out that a core component in the lens galaxy density profile could comprise an approximate internal mass sheet degeneracy (MSD), shifting the inferred value of $H_0$ without affecting the image reconstruction and without conflict with estimates of cosmological external convergence. Subsequently, TDCOSMO-IV~\cite{Birrer:2020tax} added an effective ``internal MSD" degree of freedom to their halo model fit; as a result, the error budget on $H_0$ increased to the level expected from stellar kinematics, around 10\%~\cite{Kochanek:2019ruu,Kochanek:2002rk}. Interestingly, including galaxies from the Sloan Lens ACS (SLACS) survey~\cite{Bolton:2005nf} in the kinematics analysis, and making the additional assumption that SLACS and TDCOSMO galaxies share a self-similar structure, shifted the central value of the lensing $H_0$ to the CMB value while providing some positive evidence for an  internal MSD component in the data. 
The status of the lensing $H_0$ measurements is illustrated in Fig.~\ref{fig:H0}.

In what follows we use the term ``core-MSD" instead of ``internal MSD", to highlight the fact that a natural interpretation of the added degree of freedom in the halo model corresponds to a physical core feature in the  density profile~\cite{Blum:2020mgu}.

We should emphasise that the hint~\cite{Birrer:2020tax} for a core-MSD could eventually go away after further scrutiny of uncertainties in conventional halo models~\cite{Shajib:2020ptb}. 
Nevertheless, even setting aside the results of~\cite{Birrer:2020tax}, it is interesting to examine the possibility of an actual core driving the lensing $H_0$ tension.  The question then is, what is the core made of? If the core is not traced by the light profile of the lens, then it is natural to speculate that it could come from dark matter, perhaps providing a clue to dark matter properties.

We consider the possibility that such cores come from ultralight dark matter (ULDM). ULDM has been studied extensively in recent years, and we do not give a thorough coverage of the literature here; see references to and from~\cite{Hu:2000ke,Hui2017}. ULDM is known to develop a cored density profile (``soliton") due to gravitational dynamical relaxation. The phenomenon has been identified in numerical simulations by different groups~\cite{Guzman2004,Schive:2014dra,Schive:2014hza,Schwabe:2016rze,Veltmaat:2016rxo,Mocz:2017wlg,Veltmaat:2018dfz,Levkov:2018kau,Eggemeier:2019jsu,Chen:2020cef,Schwabe:2020eac}, and is consistent with analytic considerations which show that the soliton is an energy-minimiser at fixed mass, and thus an attractor solution of the equations of motion~\cite{Chavanis:2011zi}. 

\begin{figure}[h!]
\centering
%\hspace*{-0.8cm}
 \includegraphics[scale=0.4]{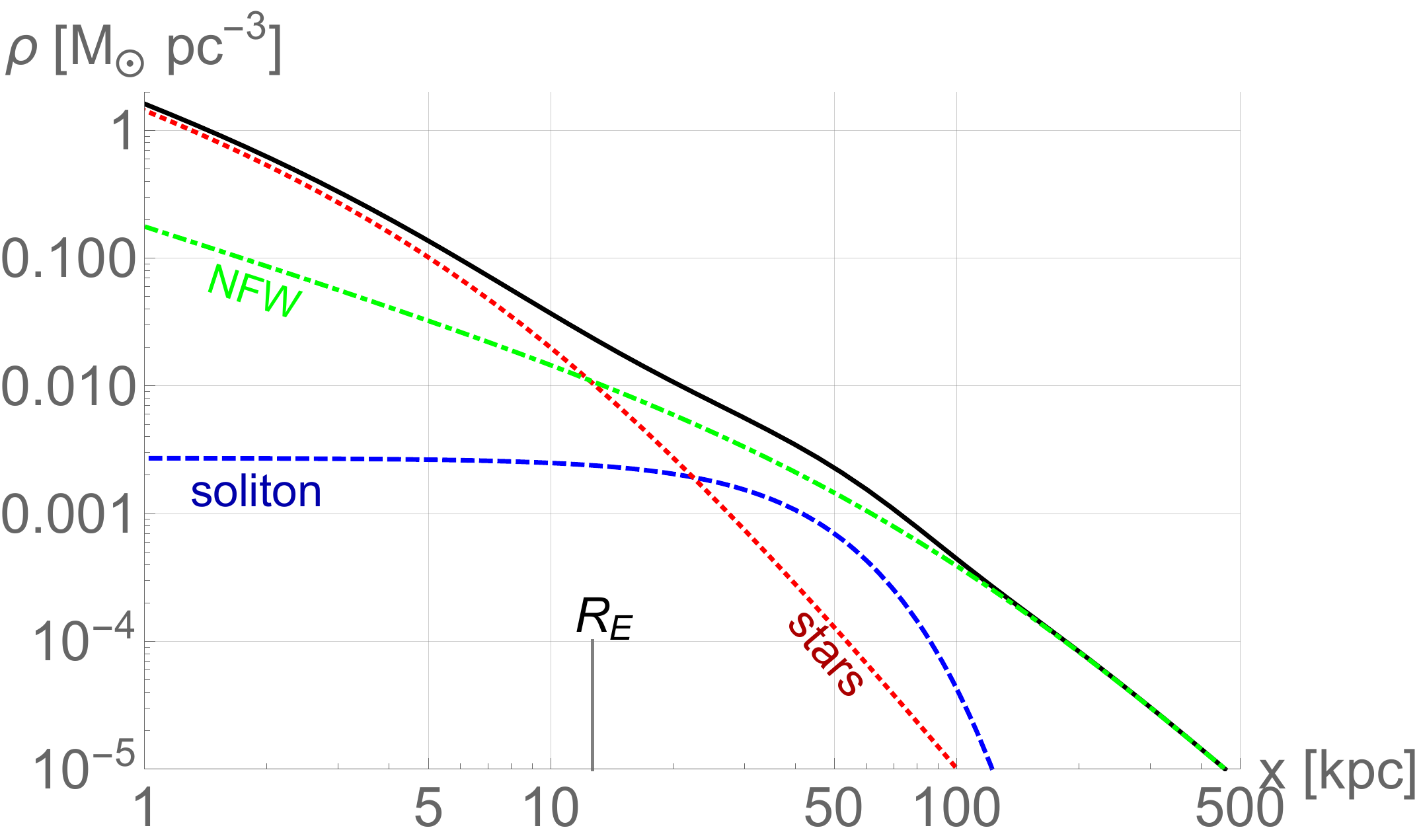}
 \caption{ULDM soliton core as a solution of the lensing $H_0$ tension. Green dash-dotted and red dotted lines show the density profiles of a cold dark matter Navarro-Frenk-White (NFW~\cite{Navarro:1996gj}) and a stellar component, respectively. The blue dashed line shows a soliton with $m=2\times10^{-25}$~eV and  $M\approx1.4\times10^{12}$~M$_\odot$, producing a shift $\delta H_0/H_0\approx0.1$. The NFW halo mass is $M_{200}\approx1.9\times10^{13}$~M$_\odot$. The halo parameters resemble TDCOSMO system DESJ0408~\cite{Millon:2019slk}. 
 }
 \label{fig:sol47}
\end{figure}

Fig.~\ref{fig:sol47} illustrates our idea. It shows the different density components (stellar mass and dark matter) of a would-be lens galaxy. We include an ULDM soliton core with a total mass of $M=1.4\times10^{12}$~M$_\odot$ at a particle mass $m=2\times10^{-25}$~eV. These ULDM parameters ($M$ and $m$) are chosen such that the core extends sufficiently far beyond the projected Einstein radius $R_\mathrm{E}$ to keep imaging errors undetectable for typical current lensing reconstruction measurement uncertainties. 

From a theoretical perspective, ULDM is a compelling possibility. If the spectrum of particles contains an ultralight boson, like the axions of some string-inspired models~\cite{Svrcek:2006yi,Marsh:2015xka}, then the phenomenon of vacuum misalignment generically predicts that such a boson would behave as dark matter if the particle mass satisfies $m\gtrsim H_0\approx10^{-33}$~eV. If the boson $\chi$ is an axion with decay constant $f$, vacuum misalignment predicts~\cite{Hui2017}
\be\label{eq:relic}\Omega_\chi&\approx&0.3\left(\frac{m}{10^{-21}~\rm eV}\right)^{\frac{1}{2}}\left(\frac{f}{10^{17}~\rm GeV}\right)^2 ,\ee
where $\Omega_\chi=\rho_\chi/\rho_{\rm crit}$ is the ratio of the ULDM relic density to the critical density of the Universe and $\Omega_\chi\approx0.3$ would saturate the total dark matter contribution $\Omega_{\rm m}$ inferred from cosmological data.  
This puts ULDM with $m\approx10^{-21}$~eV in the right order of magnitude to make up all of the dark matter if $f$ is around the grand-unification or string scale. 

As we shall see, the interesting mass range for our current analysis is actually $m\sim10^{-25}$~eV, give or take an $\mathcal{O}(1)$ factor. Cosmological and astrophysical observations imply that such ULDM can only comprise a fraction of the total dark matter. 
We thus 
define the cosmological ULDM fraction
\be\label{eq:alphax}\alpha_\chi&=&\frac{\Omega_\chi}{\Omega_\mathrm{m}};\ee
we will be led to consider $\alpha_\chi<1$. In this case, the remaining dark matter must take some other form (e.g., higher-$m$ axions).

Rotation curves of low-surface-brightness galaxies are inconsistent with $\alpha_\chi=1$ for $m\lesssim10^{-21}$~eV~\cite{Bar:2018acw}, but these constraints have not been evaluated for $\alpha_\chi<1$. 
%CMB analyses require $\alpha_\chi\lesssim0.5$ for $m\gtrsim10^{-25}$~eV~\cite{Hlozek:2017zzf}. 
Recently, Ref.~\cite{2021arXiv210407802L} reported constraints that combine galaxy clustering data~\cite{Beutler_2016} %~\cite{Beutler:2016arn} 
%\cite{10.1093/mnras/stw3298} 
with Planck15 CMB data~\cite{Ade:2015xua} (see~\cite{Hlozek:2017zzf} for an earlier analysis of the CMB data). The constraint on $\alpha_\chi$ depends on the value of $m$; for example, for $m=10^{-25}$~eV, the $2\sigma$~CL combined limit is $\alpha_\chi\lesssim0.34$, while for $m=10^{-26}$~eV the limit tightens to $\alpha_\chi\lesssim0.035$. 
Additional constraints come from the Ly-$\alpha$ forest line absorption power spectrum~\cite{Kobayashi:2017jcf}, which can be roughly summarised by $\alpha_\chi\lesssim0.16$ at $2\sigma$~CL for $m<10^{-22}$~eV. The constraint becomes weaker towards larger $m$ and disappears for $m\gtrsim10^{-20}$~eV. We note that the Ly-$\alpha$ bound of~\cite{Kobayashi:2017jcf} was not explicitely computed and must be extrapolated to the low values of $m$ where we will use it; keeping that in mind, and noting in addition that systematic uncertainties associated with the heating and ionisation history of the intergalactic medium could affect the Ly-$\alpha$ analyses to some extent, we allow ourselves to explore $\alpha_\chi$ as large as 0.2. %We will see later that the cosmological bounds are important for our scenario.  

Eq.~(\ref{eq:relic}) tells us that ULDM at $m\approx10^{-25}$~eV could easily make up $\mathcal{O}(10\%)$ of the total dark matter, in the vanilla misalignment scenario with $f\approx3\times10^{17}$~GeV. 
%It is this scenario that we can probe in the current analysis.  
%

The rest of the paper is arranged as follows.
In Sec.~\ref{s:nobar} we recap the core-MSD set-up of~\cite{Blum:2020mgu}, explaining the connection between imaging errors and the possible range of the shift in the inferred value of $H_0$. In Sec.~\ref{s:uldm} we show how an ULDM soliton produces a core-MSD profile. Using a simplified prescription to estimate imaging constraints we explore the ULDM parameter space. 
In Sec.~\ref{s:kin} we study stellar kinematics. We find that the perfect MSD limit, adopted in the kinematics analysis of TDCOSMO-IV~\cite{Birrer:2020tax}, needs to be revised if one wishes to explore a realistic physical core-MSD model. 

Our analysis suggests that ULDM could solve the lensing $H_0$ tension, provided it condenses into sufficiently massive solitons in the lens galaxies. In Sec.~\ref{s:th} we consider the theoretical consistency of this scenario. We show that ULDM solitons of roughly the right mass could indeed form naturally by dynamical relaxation. Because dynamical relaxation becomes inefficient if the cosmological ULDM fraction $\alpha_\chi$ is small, sufficiently fast soliton condensation requires that the ULDM abundance be as large as observational constraints allow it to be, $\alpha_\chi\sim0.2$. Cosmological constraints thus put some pressure on the model. Sec.~\ref{s:add} contains brief additional discussion of stellar kinematics and dynamics in well-resolved galaxies, like our own Milky Way. We summarise in Sec.~\ref{s:sum}. 

App.~\ref{a:PLan} contains technical details of the distortion of the soliton under a power-law background density profile. App.~\ref{a:PLmock} contains analyses of mock data, with references to our implementation of the ULDM model in the lensing software package~\texttt{lenstronomy}~\href{https://github.com/sibirrer/lenstronomy}{\faGithub}~\cite{Birrer:2018}. App.~\ref{a:msdkin} contains some details of the kinematics analysis.

%%%%%%%%%%%%%%%%%%%
\section{The core-MSD model}\label{s:nobar}
Consider a lensing reconstruction model $\kappa_{ 0}(\theta)$ for the convergence of the lens. A core-MSD model can be constructed from $\kappa_0(\theta)$ by adding a core component $\kappa_\mathrm{c}(\theta)$ while rescaling the original model:
\be\label{eq:msdc}\kappa(\theta)&=&\kappa_\mathrm{c}(\theta)+(1-\kappa_\mathrm{c}(\theta_\mathrm{E}))\kappa_{0}(\theta)  .\ee
Here $\theta_\mathrm{E}$ is defined by $\alpha_0(\theta_\mathrm{E})=\theta_\mathrm{E}$, where $\alpha_0(\theta)$ is the deflection angle due to $\kappa_0(\theta)$. 
At the same time, the source plane coordinates are rescaled as $\beta=(1-\kappa_\mathrm{c}(\theta_\mathrm{E}))\beta_0$. On angular scales $\theta\gg\theta_\mathrm{E}$ it is assumed that $\kappa_\mathrm{c}(\theta)\to0$ such that the core-MSD effect commutes with external convergence. 

Eq.~(\ref{eq:msdc}) is an approximate MSD if $\kappa_\mathrm{c}(\theta)$ is nearly constant up to $|\theta|$ that is sufficiently larger than $|\theta_\mathrm{E}|$.  
To be quantitative, we can define the correction $\delta_\mathrm{E}$ via:
\be\label{eq:aE}\alpha(\theta_\mathrm{E})&=&\theta_\mathrm{E}\left(1+\delta_\mathrm{E}\right),\ee
where $\alpha(\theta)$ is the deflection angle of the full model.
$\delta_\mathrm{E}$ quantifies the relative imaging error in the vicinity of $\theta\approx\theta_\mathrm{E}$, the angular range where lensing analyses have the most constraining power. 
For simplicity, in this estimate we assume the system to be spherically symmetric, so that $\alpha(\theta)=2\theta\int_0^1 \dd{z}z\kappa(z\theta)$. 
Using Eq.~(\ref{eq:msdc}) we then have
\be\label{eq:dE}\delta_\mathrm{E}&=&2\int_0^1\dd{z}z\left(\kappa_\mathrm{c}\left(z\theta_\mathrm{E}\right)-\kappa_\mathrm{c}(\theta_\mathrm{E})\right)\no\\
&=&\frac{\alpha_\mathrm{c}\left(\theta_\mathrm{E}\right)}{\theta_\mathrm{E}}-\kappa_\mathrm{c}\left(\theta_\mathrm{E}\right) .\ee
%
%If, instead, we used our previous definition, we would get:
%%
%\be\delta_\mathrm{E}&=&\frac{\alpha_c\left(\theta_\mathrm{E}\right)}{\theta_\mathrm{E}}-\kappa_\mathrm{c}\left(0\right).\ee
%%
The first line in Eq.~(\ref{eq:dE}) shows that constant $\kappa_\mathrm{c}(\theta)$ within $\theta<\theta_\mathrm{E}$ produces an MSD, and the second is convenient for quantifying corrections when $\kappa_\mathrm{c}(\theta)$ is not exactly constant. While this estimate was given for a spherical lens, it gives a good approximation of the imaging error also for the non-symmetric systems arising in realistic analyses, as we will verify using mock data. 

Consider the possibility that a lens galaxy harbours a core component, leading to a true convergence profile resembling Eq.~(\ref{eq:msdc}) with $\kappa_\mathrm{c}(\theta_\mathrm{E})>0$. In this case, both the null model $\kappa_0(\theta)$ and the core-MSD model $\kappa(\theta)$ would give a good description of the imaging data. However, the true value of $H_0$ would differ from the inferred value in the null model by:
\be\label{eq:k0dH}\frac{H_{0,\rm inferred}-H_{0,\rm true}}{H_{0,\rm true}}&\equiv&\frac{\delta H_0}{H_0}\,\approx\,\kappa_\mathrm{c}(\theta_\mathrm{E}) .\ee

Tab.~\ref{Tab:lenses} shows the values of $\delta H_0/H_0$ required to bring the different systems to accord with the CMB result. We see that $\kappa_\mathrm{c}(\theta_\mathrm{E})\approx0.1$, with some variation between systems~\cite{Suyu:2016qxx,Bonvin:2019xvn,Birrer:2018vtm,Chen:2019ejq,Wong:2019kwg}, could solve the lensing $H_0$ tension.%~\cite{Suyu:2016qxx,Bonvin:2019xvn,Birrer:2018vtm,Chen:2019ejq,Wong:2019kwg}.
\begin{table*}[htp]
\caption{Lens systems from~\cite{Millon:2019slk}. Values for $H_0$ (in km/s/Mpc) are from the PL fit  (Fig.~6 in~\cite{Millon:2019slk}). The reference ``true" $H_0$ used to define $\delta H_0/H_0$ is taken from the  CMB result $H_0 = 67.36 \pm 0.54$ km/s/Mpc~\cite{Akrami:2018vks}. $\theta_\mathrm{E}$ is in arcsec. $\sigma^\mathrm{P}$ is in km/s. On the last column we show twice the maximum relative error of the velocity anisotropy, useful for comparison with $\delta H_0/H_0$ (see discussion in Sec.~\ref{s:kin}).}
\begin{center}
\begin{tabular}{c c c c c c c c c}
\toprule
&$\delta H_0/H_0$&$\gamma $&$\theta_\mathrm{E}$&$\delta_\mathrm{E}$&$z_\mathrm{l}$&$z_\mathrm{s}$&$\sigma^\mathrm{P}$ & $ 2 |\delta\sigma^\mathrm{P}|/\sigma^\mathrm{P} $\\
\midrule
RXJ1131&$0.13^{+0.05}_{-0.06}$&$1.98$&$1.6$&$0.006$ &$0.295$&$0.654$&$320\pm20$ & $0.125$ \\
\addlinespace
PG1115&$0.23^{+0.11}_{-0.10}$&$2.18$&$1.1$&$0.02$ &$0.311$&$1.722$&$280\pm25$ & $ 0.178 $\\
\addlinespace
HE0435&$0.06^{+0.07}_{-0.07}$&$1.87$&$1.2$&$0.025$&$0.4546$&$1.693$&$220\pm15$ & $0.136$ \\
\addlinespace
DESJ0408&$0.11^{+0.04}_{-0.04}$&$2$&$1.9$&$0.01$&$0.597$&$2.375$&$230\pm27$ & $ 0.235 $\\
\addlinespace
WFI2033&$0.08^{+0.05}_{-0.04}$&$1.95$&$0.9$&$0.016$ &$0.6575$&$1.662$&$250\pm19$ & $ 0.152 $\\
\addlinespace
J1206&$-0.01^{+0.08}_{-0.07}$&$1.95$&$1.2$&$0.025$ &$0.745$&$1.789$&$290\pm30$ & $ 0.207 $\\
\bottomrule
\end{tabular}
\end{center}
\label{Tab:lenses}
\end{table*}%
%

%%%%%%%%%%%%%%%%%%%
\section{Core-MSD with an ULDM soliton}\label{s:uldm}
An ULDM soliton could produce the $\kappa_\mathrm{c}$ term in Eq.~(\ref{eq:msdc}). We now derive some results that are useful for the lensing analysis; for a detailed discussion and more references concerning ULDM solitons, we refer the reader to~\cite{Bar:2018acw}. 

The ULDM soliton field is described by a function $\chi(r)$, where we define the rescaled coordinate $r=mx$. The mass density associated with $\chi$ is given by
\be\label{eq:solrho}\rho&=&\frac{m^2}{4\pi G}\chi^2 , \ee
where $G$ is Newton's constant. 
The field $\chi$ and the Newtonian gravitational potential sourced by it, $\Phi$, satisfy the Schrodinger-Poisson equations (SPE)~\cite{Bar:2018acw}
\be\label{eq:SPE1} \partial^2_r(r\chi)&=&2r\left(\Phi+\Phi_{\rm ext}-\tilde\gamma\right)\chi  ,\\
\label{eq:SPE12}\partial^2_r(r\Phi)&=&r\chi^2  .
\ee
We include a background gravitational potential $\Phi_{\rm ext}$, coming from stars and from other (non-ULDM) contributions to the DM. Indeed, in the problem at hand the soliton contributes just a small part to the mass density of the lens near the Einstein radius, so we  anticipate typically $|\Phi_{\rm ext}|>|\Phi|$. The variable $\tilde\gamma$ is an eigenvalue that characterises the solution. We are interested in the lowest-energy solution, where $\chi$ starts off constant at $r\to0$ and decays to zero with no nodes. We solve the SPE numerically. 

The solution is fixed by a single parameter that we can take to be the value of $\chi$ at $r=0$. We thus define the solution $\chi_\lambda(r)$ via
\be\label{eq:lamdef}\chi_\lambda(r=0)&=&\lambda^2  ,\ee
with a real parameter $\lambda$. 
It is convenient to use the scaling relation~\cite{Bar:2018acw}
 \be\label{eq:chilam0}\chi_\lambda(r;\Phi_{\rm ext}(r))&=&\lambda^2\chi_1(\lambda r;\lambda^{-2}\Phi_{\rm ext}(\lambda^{-1}r))  ,\ee
meaning that in numerical investigations, it is always enough to compute $\chi_1$. For clarity, in Eq.~(\ref{eq:chilam0}) we explicitly note how the external potential enters the solution. 

It is convenient to introduce an approximation with which properties of the soliton can be derived analytically. We choose
\be\label{eq:abapp}\chi_{1}(r)&\approx&\frac{1}{\left(1+a^2r^2\right)^b}  ,\ee
where the coefficients $a$ and $b$ are fitted numerically to the exact solution. 
For a self-gravitating soliton (the limit $\Phi_{\rm ext}\to0$), we obtain $a\approx0.23$ and $b\approx3.9$. When $\Phi_{\rm ext}\neq0$, the coefficients $a$ and $b$ depend on $\lambda$ and $\Phi_{\rm ext}$ via the combination $\lambda^{-2}\Phi_{\rm ext}(\lambda^{-1}r)$.

With the approximation of Eq.~(\ref{eq:abapp}), the soliton mass is
\be\label{eq:Mlam}
M_{\lambda}&=&\frac{\lambda}{Gm}\int \dd{r}r^2\chi^2_1(r)\approx\frac{\lambda}{Gm}\frac{\sqrt{\pi}}{a^3}\frac{\Gamma\left(2b-\frac{3}{2}\right)}{4\Gamma\left(2b\right)}  . 
\ee
The convergence, deflection angle, and lensing potential are:
\be\label{eq:kaplam}
\kappa_{\lambda}(\theta)&\approx&\frac{\lambda^3m}{4\pi G\Sigma_\mathrm{c}}\frac{\sqrt{\pi}}{a}\frac{\Gamma\left(2b-\frac{1}{2}\right)}{\Gamma\left(2b\right)}\frac{1}{\left(1+\frac{\theta^2}{\theta_\mathrm{c}^2}\right)^{2b-\frac{1}{2}}} ,\\
\label{eq:alam}\alpha_{\lambda}(\theta)&\approx&\kappa_\lambda(0)\frac{2\theta_\mathrm{c}^2}{(4b-3)\theta}\left(1-\left(1+\frac{\theta^2}{\theta_\mathrm{c}^2}\right)^{\frac{3}{2}-2b}\right)  ,\\
\label{eq:psilam}\psi_\lambda(\theta)&\approx&\kappa_\lambda(0)\frac{\theta^2}{2}\,_3\mathcal{F}_2\left[\left\{1,1,2b-\frac{1}{2}\right\},\{2,2\};-\frac{\theta^2}{\theta_\mathrm{c}^2}\right]  ,\no\\&&\ee
where we defined the core angle
\be\label{eq:thc}\theta_\mathrm{c}&=&\frac{1}{\lambda amD_\mathrm{l}} .\ee
The critical density entering the convergence computation is $\Sigma_\mathrm{c}^{-1}=4\pi G D_\mathrm{l}D_\mathrm{ls}/D_\mathrm{s}$, where $D_{\rm l,s,ls}$ are the angular diameter distances to the lens, source, and between the lens and the source. $_p\mathcal{F}_q\left[\vec a,\vec b;z\right]$ is the generalised hypergeometric function\footnote{A rapidly converging expression is $_3\mathcal{F}_2\left[\left\{1,1,2b-\frac{1}{2}\right\},\{2,2\};z\right]=\frac{2}{z(3-4b)}\left(\log(1-z)-\sum_{n=1}^\infty\frac{\Gamma\left(\frac{5}{2}-2b+n\right)}{\Gamma\left(\frac{5}{2}-2b\right)}\frac{z^n}{(z-1)^n\,n\,n!}\right)$.}. 
%In these results one should use $a=a(A/\lambda^2),\;b=b(A/\lambda^2)$. With the same understanding for $a$ and $b$, and u
In the MSD limit, $\theta_\mathrm{c}\gg\theta$, one can verify that $\kappa_\lambda(\theta)\approx\kappa_\lambda(0)$, $\alpha_\lambda(\theta)\approx\kappa_\lambda(0)\theta$, and $\psi_\lambda\approx\kappa_\lambda(0)\theta^2/2$. 

Adopting the soliton as our core-MSD component, we set $\kappa_\mathrm{c}(\theta)\equiv\kappa_\lambda(\theta)$ in Eq.~(\ref{eq:msdc}). 
From Eqs.~(\ref{eq:dE}) and~(\ref{eq:k0dH}) we get
\be\label{dH2Hmsd}\frac{\delta H_0}{H_0}&\approx&\frac{\kappa_\lambda(0)}{\left(1+\frac{\theta_\mathrm{E}^2}{\theta_\mathrm{c}^2}\right)^{2b-\frac{1}{2}}} \ee
and
\be\label{eq:dElam}\delta_\mathrm{E}&\approx&\frac{\left(1+\frac{\theta_\mathrm{E}^2}{\theta_\mathrm{c}^2}\right)^{2b-\frac{1}{2}}-\left(2b-\frac{1}{2}\right)\frac{\theta_\mathrm{E}^2}{\theta_\mathrm{c}^2}-1}{\left(2b-\frac{3}{2}\right)\frac{\theta_\mathrm{E}^2}{\theta_\mathrm{c}^2}}\frac{\delta H_0}{H_0}  .\ee
For $\theta_\mathrm{E}\ll\theta_\mathrm{c}$ we have $\delta_\mathrm{E}\approx\left(b-\frac{1}{4}\right)\frac{\theta_\mathrm{E}^2}{\theta_\mathrm{c}^2}\frac{\delta H_0}{H_0}$. This shows how imaging uncertainties, roughly summarised by $\delta_{\rm E}$, constrain the shift $\delta H_0/H_0$ at given soliton core angle $\theta_{\rm c}$.

In App.~\ref{a:PLan} we calculate how the soliton profile is distorted in the presence of a power-law (PL) background. We find that Eq.~(\ref{eq:abapp}) is still a good approximation, sufficient for our needs; the effect of the background density is to modify the numerical values of the coefficients $a$ and $b$.

Before proceeding to observational constraints, we comment on the parameter space of the model. As noted in the beginning of this section, at a fixed value of the ULDM particle mass $m$, the soliton is a single-parameter function. While the scaling parameter $\lambda$ from Eq.~(\ref{eq:lamdef}) is convenient for analytical expressions, in making contact with observations we prefer to use the total soliton mass $M$, substituting $\lambda\to\lambda(M,m)$ using Eq.~(\ref{eq:Mlam}). (The detailed matching, but not the basic procedure, is slightly modified with a background potential as explained in App.~\ref{a:PLan}.) All other properties of the core (the convergence, for example) then depend only on $m$ and $M$. The full parameter space is therefore covered when we analyse our results in terms of $m$ and $M$. 

%As a last comment, we note that our results can be easily extended to cover 
%The core toy model in~\cite{Birrer:2020tax} was different from an ULDM soliton, but it too can be matched to (the square of) Eq.~(\ref{eq:abapp}). Eqs.~(\ref{eq:s2c}-\ref{eq:f2c}) hold, with $b=3/4$ for the toy model replacing $b\approx3.9$ for a soliton, and with the same $\theta_{\rm c}$ parameter. We approximately convert between the effective $\theta_{\rm c}$ parameters by matching the convergence of the two models at $\theta_{\rm E}$.

\subsection{Constraints on ULDM from TDCOSMO systems}\label{ss:dat}
\begin{figure*}
\centering
%\hspace*{-0.8cm}
 \includegraphics[scale=0.46]{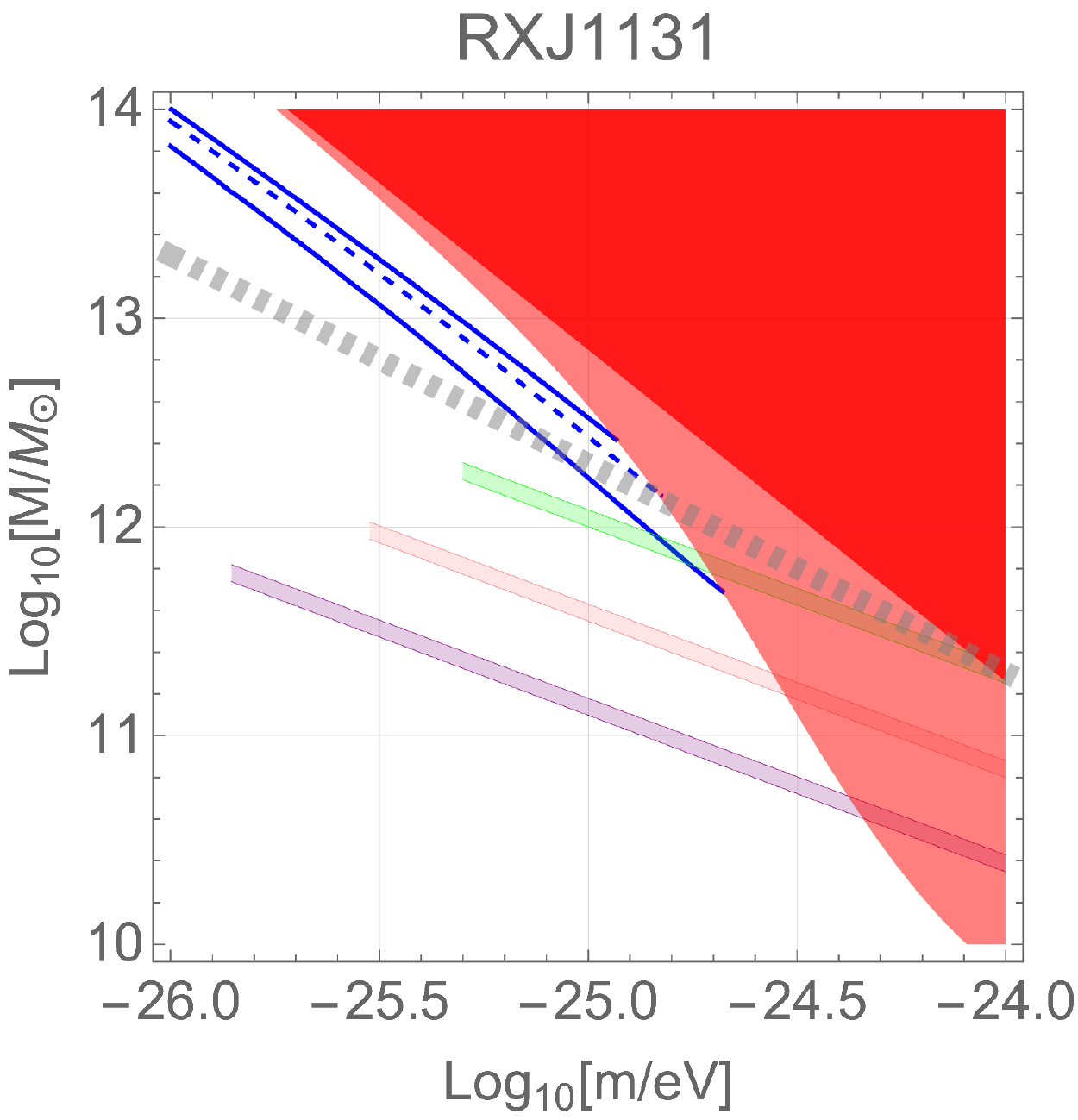}
 \includegraphics[scale=0.46]{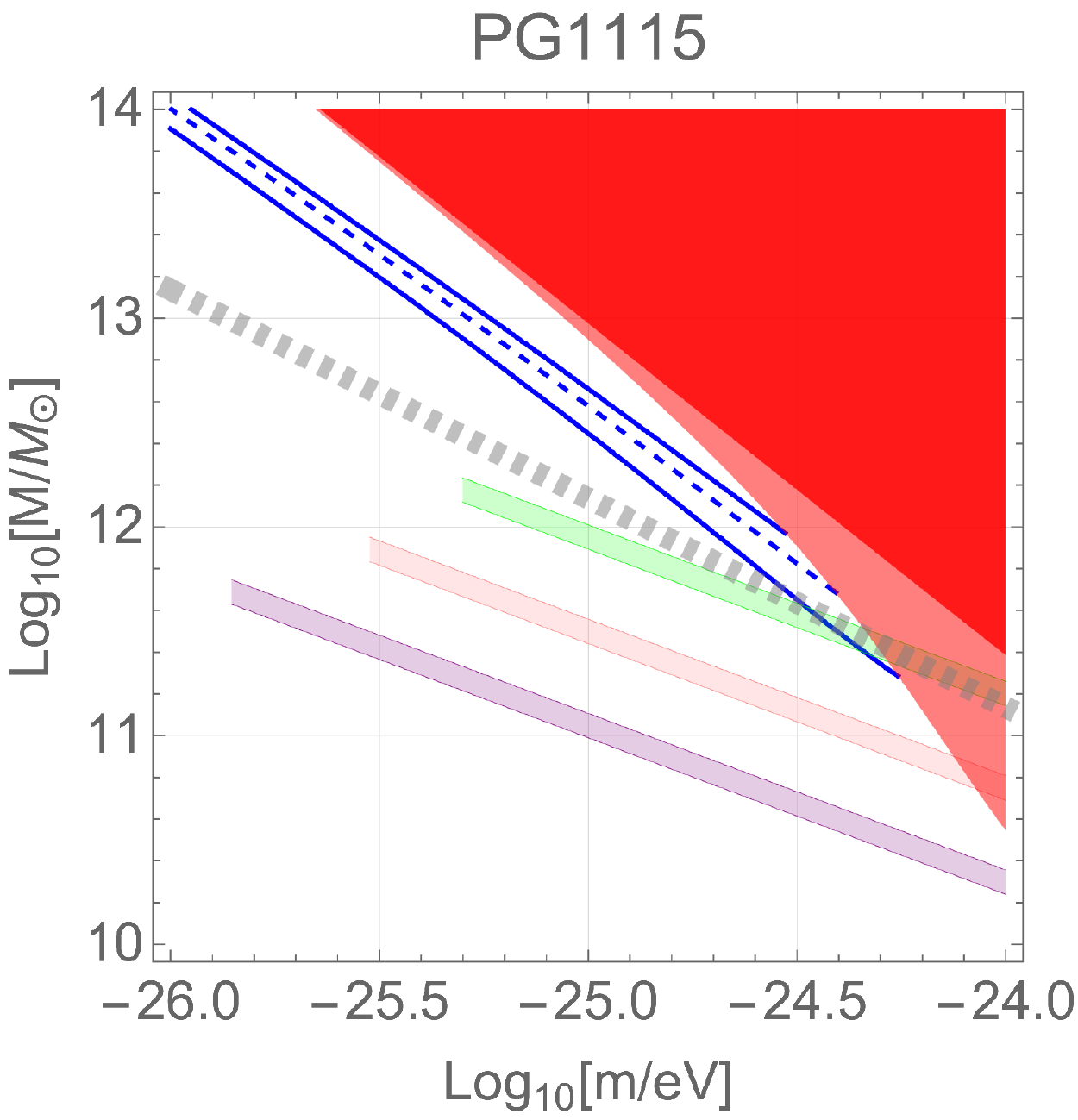}
 \includegraphics[scale=0.46]{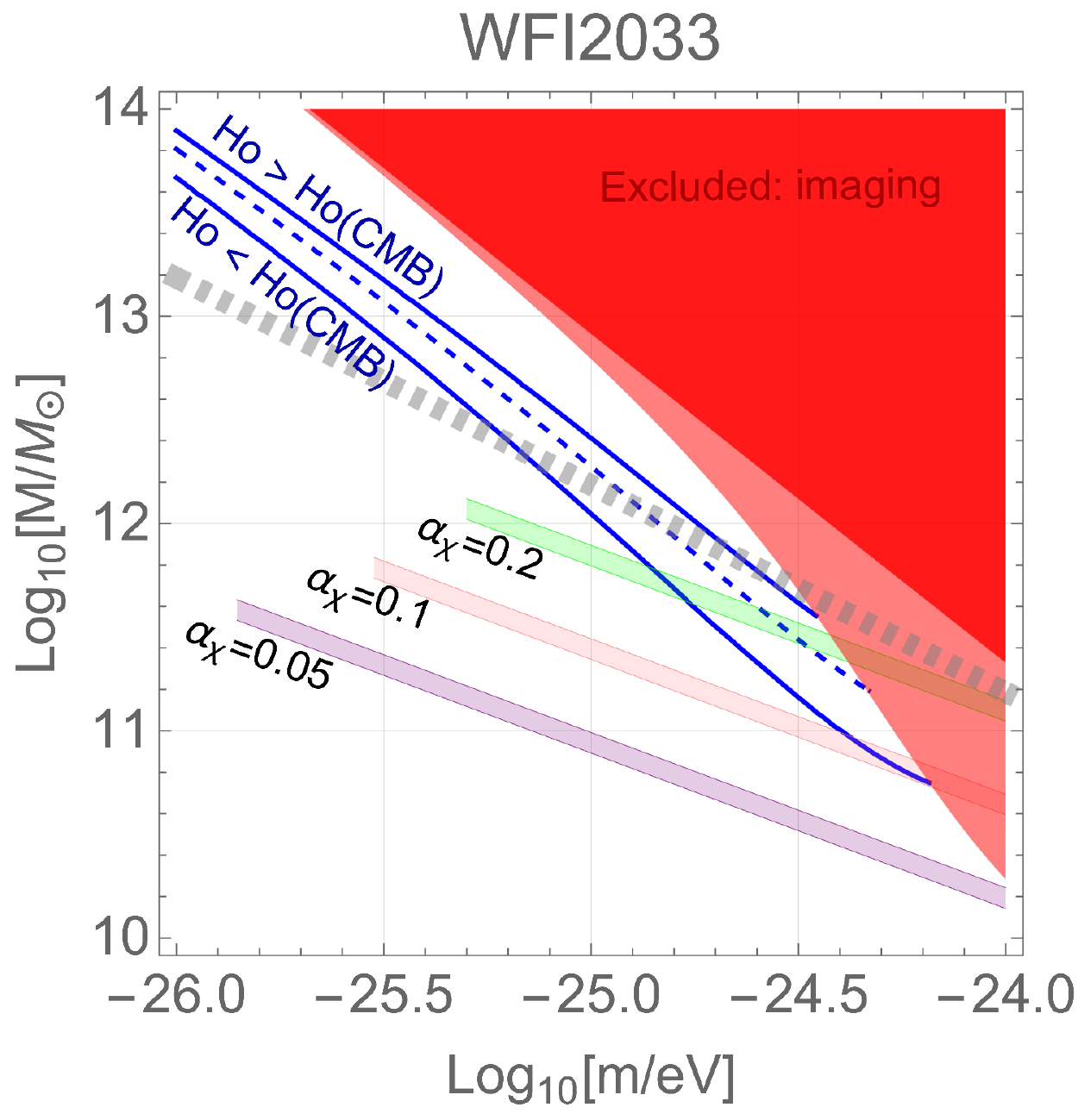}
 \includegraphics[scale=0.46]{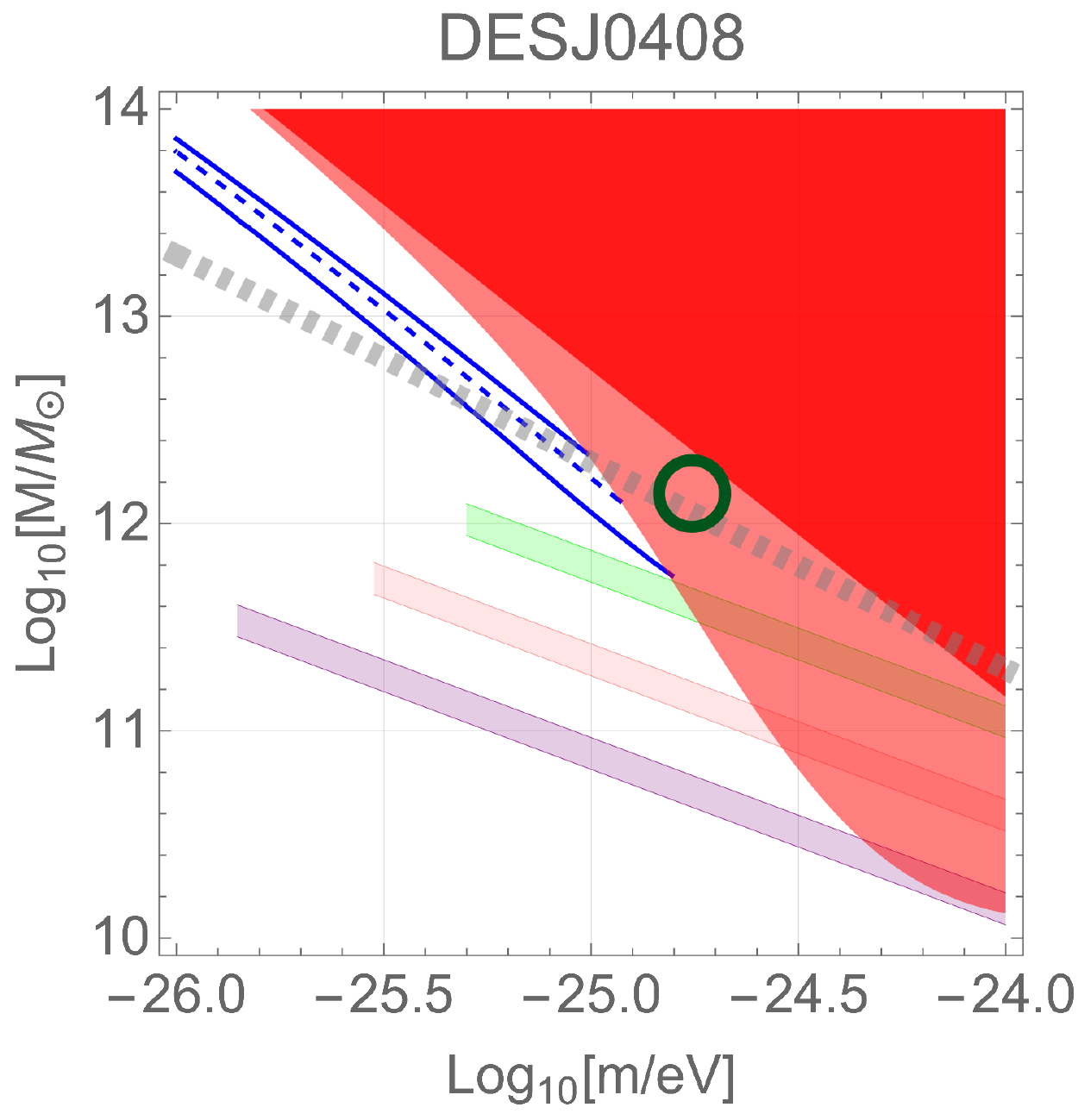}
 \includegraphics[scale=0.46]{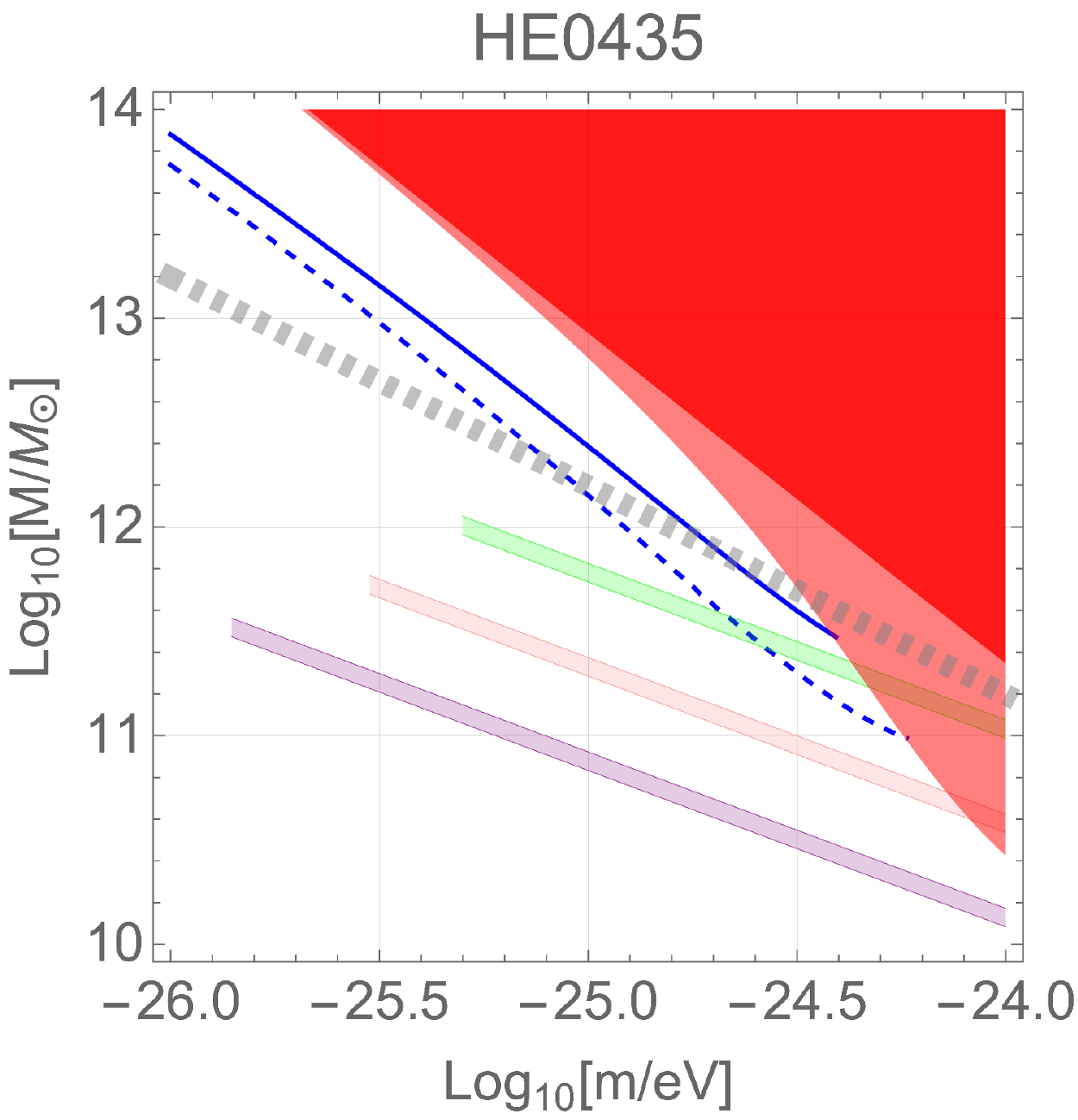}
 \includegraphics[scale=0.46]{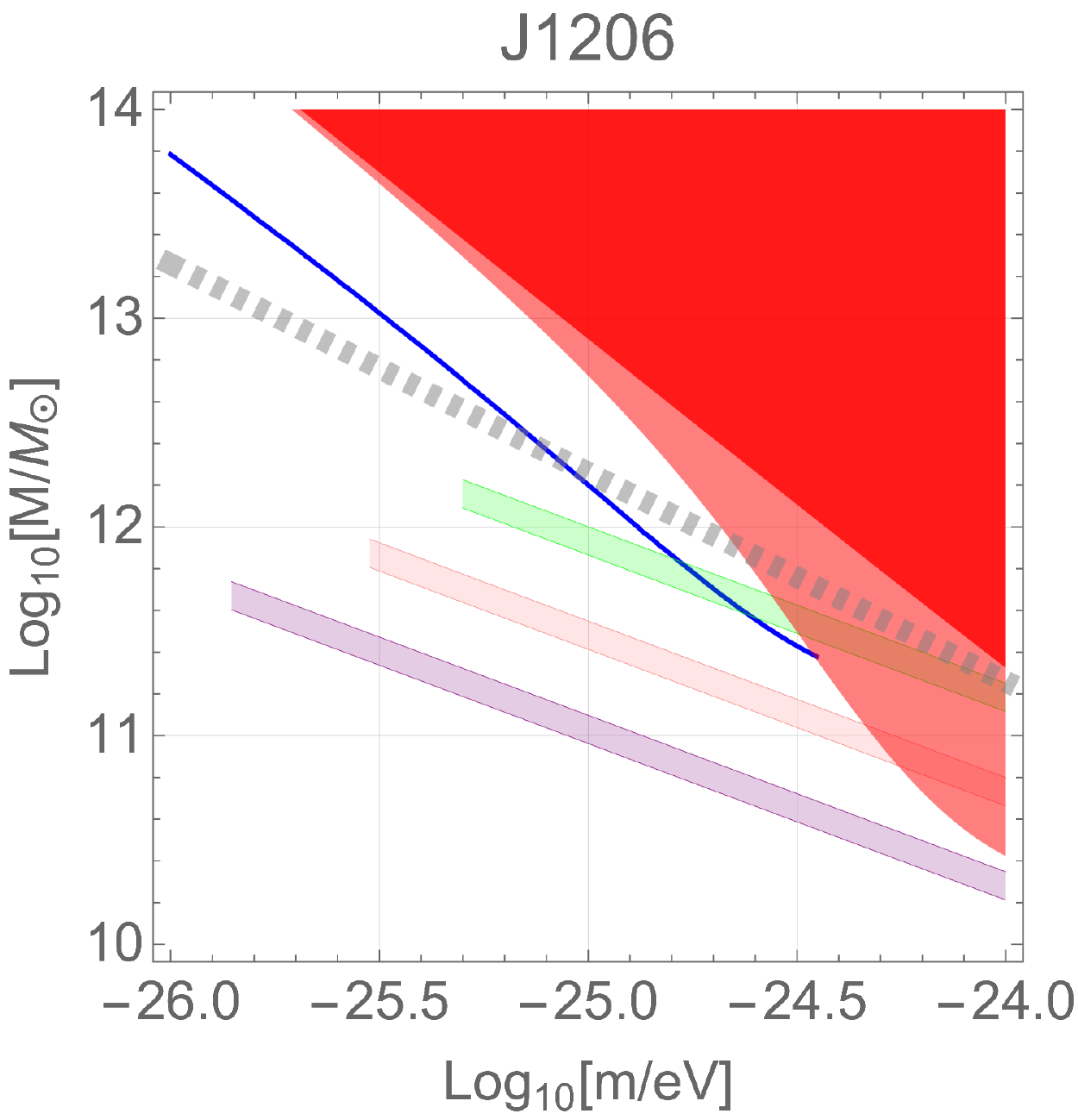}
 \caption{ULDM soliton as a solution of the lensing $H_0$ tension. Panels correspond to different systems in Tab.~\ref{Tab:lenses}. The dark red shaded region shows where $\delta_\mathrm{E}$ exceeds its limits from Tab.~\ref{Tab:lenses}, computed neglecting the effect of an external mass distribution on the soliton. The pale shaded region shows the same constraint, including the effect of an external PL potential. Along the blue dashed line, $\delta H_0/H_0$ matches the value from Tab.~\ref{Tab:lenses}, with solid lines delimiting the uncertainty; this result also includes the PL background. (In J1206 and HE0435 the central value and/or lower limit of $\delta H_0/H_0$ are compatible with zero.) 
The relaxation estimate of Eq.~(\ref{eq:Mmaxdat}) (see Sec.~\ref{s:th}) is shown by green, pink, and purple bands for $\alpha_\chi=0.2,\,0.1,\,0.05$, respectively, using $\sigma_0=\sigma^\mathrm{P}$. The band width is defined by the uncertainty in $\sigma^\mathrm{P}$. We truncate each constant-$\alpha_{\chi}$ band at small $m$ according to the cosmological constraints from Ref.~\cite{2021arXiv210407802L}. 
 %The age of each lens galaxy [$t_{\rm gal}$ in Eq.~(\ref{eq:Mmaxdat})] is taken as the FRW time between $z=20$ and the lens redshift $z_\mathrm{l}$. 
 The saturation estimate [$(K/M)_\lambda<1.5(A/2)$] is shown by thick dashed grey line. The circle in the panel of DESJ0408 marks the set-up of Fig.~\ref{fig:sol47}.}
 \label{fig:sys}
\end{figure*}

We are ready for a rough assessment of the lensing $H_0$ tension in the ULDM model. 
In Fig.~\ref{fig:sys} we explore $\delta H_0/H_0$ and $\delta_\mathrm{E}$ as function of the ULDM particle mass $m$ ($x$-axis) and soliton mass $M$ ($y$-axis). The different panels correspond to the different systems in Tab.~\ref{Tab:lenses}. The information in the plot is as follows. 

We begin with results that include the effect of a background (non-ULDM) external potential, modelled as a pure PL, using the results in App.~\ref{a:PLan}. For simplicity, the same PL index $\gamma=2$ is used for all systems, but the value of $\theta_\mathrm{E}$ and $\delta_{\rm E}$ for each system is as in Tab.~\ref{Tab:lenses}. In the pale red shaded region, $\delta_\mathrm{E}$ exceeds its corresponding value from Tab.~\ref{Tab:lenses}. This region is disfavoured by the imaging data. Along the blue dashed line, $\delta H_0/H_0$ matches the value required to solve the $H_0$ tension. The solid blue lines delimit the uncertainty on $\delta H_0/H_0$ for each system. (Other curves in Fig.~\ref{fig:sys} correspond to theoretical constraints and are explained in Sec.~\ref{s:th}.)

We also show how the imaging constraints change if the external PL density is not included in the soliton  computation. The result is shown by the dark red shaded region. 
The imaging constraint is generally weaker when the PL effect is not included, compared to when it is (i.e., the dark shaded region is contained inside the pale region), because the background potential causes the soliton to contract inwards at fixed $m$ and $M$, decreasing $\theta_\mathrm{c}$ and leading to stronger violation of the MSD. %This tightens the $\delta_\mathrm{E}$ constraint compared to the result that would obtain neglecting the background. 

The fact that the PL background analysis provides stronger imaging constraints, compared with the self-gravity case, illustrates the sensitivity of the analysis to the detailed mass profile of the lens. However, the soliton contraction is mostly driven by the cuspy PL mass distribution at small $r\ll R_\mathrm{E}$, where the lensing observables are not well constrained. In fact, the observed stellar surface brightness of the lenses display cores rather than cusps on distances $r\ll R_\mathrm{E}$, where the stellar density dominates over the DM. As a result, physically-motivated composite stellar+DM halo models, adjusted to fit the stellar light profile, predict a contraction effect on the soliton that is less significant than in the PL background. The imaging constraint in these more realistic background models are closer to the self-gravitating soliton result.

In Fig.~\ref{fig:sol47} we show a soliton solution of the lensing $H_0$ tension, using a composite stellar+DM model that mimics the properties of the system DESJ0408. The solution has $m=2\times10^{-25}$~eV and $M\approx1.4\times10^{12}$~M$_\odot$, marked in Fig.~\ref{fig:sys} by a circle (bottom-left panel). This solution has $\delta H_0/H_0\approx0.1$ and we have verified that it is compatible with the requirement $|\delta_\mathrm{E}|<0.01$, valid for this system. The fact that this solution would seem to be excluded in the PL background analysis is due to the exaggerated soliton contraction in the PL case. %The realistic constraint is more consistent with the self-gravitating solution.

We can use the self-gravitating soliton case to understand the imaging constraints parametrically.  
In this limit $a\approx0.23$ and $b\approx3.9$ can be used in Eqs.~(\ref{eq:Mlam}-\ref{eq:dElam}) and the  $H_0$ shift is
\be\label{eq:H0PLapp}\frac{\delta H_0}{H_0}&\approx&2.9\frac{\lambda^3m}{4\pi G\Sigma_\mathrm{c}} \ .
\ee
On the other hand, in the same self-gravity limit we have $\delta_\mathrm{E}\approx3.6\left(\theta_\mathrm{E}/\theta_\mathrm{c}\right)^2\delta H_0/H_0$. Demanding $\delta_\mathrm{E}\lesssim0.01$, as in typical systems, and setting $\delta H_0/H_0\approx0.1$, we should impose $\theta_\mathrm{c}\gtrsim6\theta_\mathrm{E}$, or
\be\label{eq:lm0}\lambda m&\lesssim&\frac{0.7}{D_\mathrm{l}\theta_\mathrm{E}} \ .\ee
Combining Eqs.~(\ref{eq:H0PLapp}) and~(\ref{eq:lm0}) we obtain,
\be&&{\rm self-gravity\;approximation:}\no\\
\label{eq:m0} m&\lesssim&%\left(\frac{2.9}{0.1\times4\pi G\Sigma_\mathrm{c}}\right)^{\frac{1}{2}}\left(\frac{0.7}{D_\mathrm{l}\theta_\mathrm{E}}\right)^{\frac{3}{2}}\left(\frac{\delta H_0/H_0}{0.1}\right)^{-\frac{1}{2}}\\
%
%&\approx&
10^{-24}~{\rm eV}\left(\frac{D_\mathrm{ls}}{D_{s}}\right)^{\frac{1}{2}}\left(\frac{1''}{\theta_\mathrm{E}}\right)^{\frac{3}{2}}\left(\frac{1~\rm Gpc}{D_\mathrm{l}}\right)\left(\frac{0.1}{\delta H_0/H_0}\right)^{\frac{1}{2}}.\no\\&&\ee
Again, the presence of an external potential (PL or composite) contracts the soliton inward to some extent at fixed $M$ and $m$, shifting the upper limit of Eq.~(\ref{eq:m0}) to somewhat lower $m$. 

Our discussion of the imaging constraints was simplistic, in that we used the rough Einstein angle criterion $\delta_{\rm E}$ to constrain the possible shift to $H_0$. In comparison, the likelihood function in real lensing analyses contains detailed extended source information as well as multiple modelling parameters, experimental seeing limitations, etc. 
In App.~\ref{a:PLmock} we present a numerical study of a mock system, including most of these complications, using~\texttt{lenstronomy}. This numerical study serves two purposes. First, we introduce an implementation of the ULDM module in~\texttt{lenstronomy}. In future work we plan to use this tool to test the ULDM model including the full lensing likelihood. Second, this exercise allows us to test the accuracy of the simple $\delta_{\rm E}$ criterion. We find that the naive $\delta_{\rm E}$ criterion is slightly conservative compared with a full analysis: for example, at fixed $\delta H_0/H_0\approx0.1$, we find that a full numerical analysis yields a constraint on $\theta_{\rm c}$ (and therefore, equivalently, on $m$ at fixed $M$) that is about a factor of 2 weaker than the constraint we would obtain using the naive $\delta_{\rm E}$ criterion.

\section{Stellar kinematics}\label{s:kin}
Stellar kinematics measurements break the MSD, and are the limiting observational factor to a core-MSD shift of  $H_0$. The basic observable is the luminosity-weighted velocity dispersion along the line of sight, $ \sigma_\mathrm{los} $, given by~\cite{Mamon:2005}
\be \label{sigmaFull_Mamon05}
\sigma^2_\mathrm{los}(\theta) &=& \frac{2 G}{I(\theta)} \int_{1}^{\infty} \frac{\dd{y}}{y} K \qty(y, \frac{\theta_\mathrm{a}}{\theta}) l\left(yD_{\rm l}\theta\right) M\left(yD_{\rm l}\theta\right).\no\\&&
\ee
Here, $ l(r) $ is the stellar luminosity density,  $I(\theta)$ is the surface brightness, $M(r)$ is the total enclosed mass, and the function $ K(u,u_a) $ encodes the velocity anisotropy profile with Osipkov-Merritt~\cite{Osikpov:1979,Merritt:1985} anisotropy radius $r_a=D_{\rm l}\theta_a$~\cite{Mamon:2005}. 
For analytical estimates, we note that the isotropic velocity limit gives
$K(u, \infty) =  \sqrt{ 1 - 1/u^2}$.

The core-MSD model enters Eq.~\eqref{sigmaFull_Mamon05} via the mass profile, $M(r)=M_{\rm c}(r)+(1-\kappa_{\rm c})M_0(r)$, where $M_0(r)$ comes from the null model and $M_{\rm c}(r)$ from the core. 
The dispersion of the full model is related to that of the null model, $\sigma_{{\rm los},0}^2$, via
\be\label{eq:s2s20app}
\frac{\sigma_{\rm los}^2}{\sigma_{{\rm los},0}^2} &=& 1 - \kappa_\mathrm{c}\left(1 - \delta_{\rm c}\right),\\
\delta_{\rm c}&=&\frac{1}{\kappa_{\rm c}}\frac{\sigma^2_\mathrm{c}}{\sigma_{{\rm los},0}^2},
\ee
where $\sigma_{\rm c}^2$ is the velocity dispersion due to the core itself. In general, all of $\sigma_{{\rm los},0}^2,\,\sigma_{\rm c}^2,\,\sigma_{\rm los}^2,$ and $\delta_{\rm c}$ depend on the measurement point $\theta$. 
In Eq.~(\ref{eq:s2s20app}), the term $\delta_{\rm c}$ parametrises the deviation from the perfect MSD limit. It becomes small for $\theta_{\rm c}\gg\theta$, but may be quantitatively relevant once we consider a finite soliton core, and once kinematics data probing $\theta$ not much smaller than $\theta_c$ is used. 

To see this, consider an isothermal PL profile for the null model, for which $ M_0(r) = 2 \sigma^2_\mathrm{v} r / G$ where $\sigma_{\rm v}^2$ is the physical velocity dispersion. In convenient angular variables we can trade $\sigma^2_{\rm v}$ for $\theta_{\rm E}$, noting that $M_0(\theta)=2\Sigma_{\rm c}D_{\rm l}^2\theta_{\rm E}\theta$. We also take the isotropic limit of $K$, and consider the Hernquist profile~\cite{Hernquist:1990} for the luminosity density, 
%\footnote{The surface brightness is then \be I(\theta)&=&\frac{l_0D_{\rm l}\theta_*}{4\pi}\frac{\left(2+\frac{\theta^2}{\theta_*^2}\right)g\left(\frac{\theta}{\theta_*}\right)-3\sqrt{\left|1-\frac{\theta^2}{\theta_*^2}\right|}}{\left|1-\frac{\theta^2}{\theta_*^2}\right|^{\frac{5}{2}}},\ee where $g(y)={\rm atanh}\left(\sqrt{1-y^2}\right)$ for $y<1$ and $g(y)={\rm atan}\left(\sqrt{y^2-1}\right)$ for $y\geq1$.}
$l(r)= l_0r_*^4/\left[2\pi r(r+r_*)^3\right]$. The parameter $r_*$ is related to the commonly used effective radius $r_{e}$ via $r_e\approx1.8r_*$~\cite{Hernquist:1990}. With these simplifications, and using Eq.~(\ref{eq:abapp}) for the soliton profile (with $b\approx3.9$), we obtain
\be\label{eq:s2c}
\delta_{\rm c}(\theta) 
 &=& 
\frac{\pi\theta^2}{3 \theta_{\rm E}\theta_{\rm c}} f\left(\frac{\theta_{\rm c}}{\theta},\frac{\theta_*}{\theta}\right) \\
&\approx&0.31\left(\frac{10''}{\theta_{\rm c}}\right)\left(\frac{1''}{\theta_{\rm E}}\right)\left(\frac{\theta}{0.5''}\right)^2\frac{f\left(\frac{\theta_{\rm c}}{\theta},\frac{\theta_*}{\theta}\right)}{f\left(20,0.5\right)},\no
\ee
where
\be\label{eq:f2c} f(y_{\rm c},y_*)&=&\frac{\frac{2\Gamma\left(2b\right)}{\sqrt{\pi}\Gamma\left(2b-\frac{1}{2}\right)}\int_{1}^{\infty} \dd{y} \frac{\sqrt{{y}^2 - 1}}{\left(y + y_*\right)^{3}}\,  _2F_1\left(\frac{3}{2},2b,\frac{5}{2},-\frac{y^2}{y_c^2}\right)}{ \int_{1}^{\infty}\dd{y}  \frac{\sqrt{y^2 - 1}} {y^2\left(y + y_*\right)^{3}}   }.\no\\&&\ee

TDCOSMO-IV~\cite{Birrer:2020tax}, in considering the effect of an ``internal MSD", have assumed in practice the perfect MSD limit $\delta_{\rm c}=0$ in their kinematics analysis of TDCOSMO and SLACS systems. The approximation was tested using a mock system with\footnote{We thank S. Birrer for clarifications about this point.} $\theta_{\rm e}=0.2''$ and several core radii\footnote{The core toy model in~\cite{Birrer:2020tax} was different from our soliton core. The kinematics effect is approximately matched between the two models for $\theta_{\rm c}^{\rm (toy)}\approx0.5\theta_{\rm c}^{\rm (soliton)}$. We give some more details on this comparison in App.~\ref{a:msdkin}.}. 
However, the parametric breaking of the MSD, captured by Eq.~(\ref{eq:s2c}), was not explored for different values of $\theta_{\rm e}$ or the baseline $\theta_{\rm E}$ and system redshifts (equivalently $\sigma_{\rm v}$). As Eq.~(\ref{eq:s2c}) suggests a strong dependence on the  kinematics observation point, $\theta$, it is important to check to what extent the MSD limit is expected to hold across different systems. 

For TDCOSMO systems~\cite{Millon:2019slk,Birrer:2020tax}, the kinematics constraints were based on a single effective measurement centred on $\theta=0$ and averaged over an aperture $\mathcal{A}$, weighted by the surface brightness $I(\theta)$. To be more precise, the observationally accessible dispersion $\sigma^\mathrm{P}$ is given by~\cite{Suyu:2009by,Birrer:2015fsm} 
\be
(\sigma^\mathrm{P})^2  &=& \frac{\int_\mathcal{A}\dd^2\theta [I(\theta) \sigma^2_\mathrm{los}(\theta) * \mathcal{P}]}{ \int_\mathcal{A}\dd^2\theta [I(\theta) * \mathcal{P}]},
\ee
where $ \mathcal{P} $ is the seeing. 
It is natural to define 
\be
\frac{\left(\sigma^{\rm P}\right)^2}{\left(\sigma_0^{\rm P}\right)^2}&=&1-\kappa_c\left(1-\Delta_{\rm c}\right),\label{eq:Dc1}\\
\Delta_{\rm c}&=&\frac{\int_{\mathcal{A}}\dd\theta\theta I(\theta)\sigma_{\rm los,0}^2(\theta)\delta_{\rm c}(\theta)*\mathcal{P}}{\int_{\mathcal{A}}\dd\theta\theta I(\theta)\sigma_{\rm los,0}^2(\theta)*\mathcal{P}}.\label{eq:Dc2}
\ee
From this expression and the previously quoted results, the correction term $\Delta_{\rm c}$ can be evaluated numerically. It depends on $\theta_{\rm c}$, $\theta_*$ (equivalently $\theta_{\rm e}$), the aperture $\mathcal{A}$, and the seeing $\mathcal{P}$. The main point to explore is how $\Delta_{\rm c}$ reacts to different values of $\theta_{\rm e}$ and $\theta_{\rm c}$.  

\begin{figure}
\centering
%\hspace*{-0.8cm}
 \includegraphics[scale=0.45]{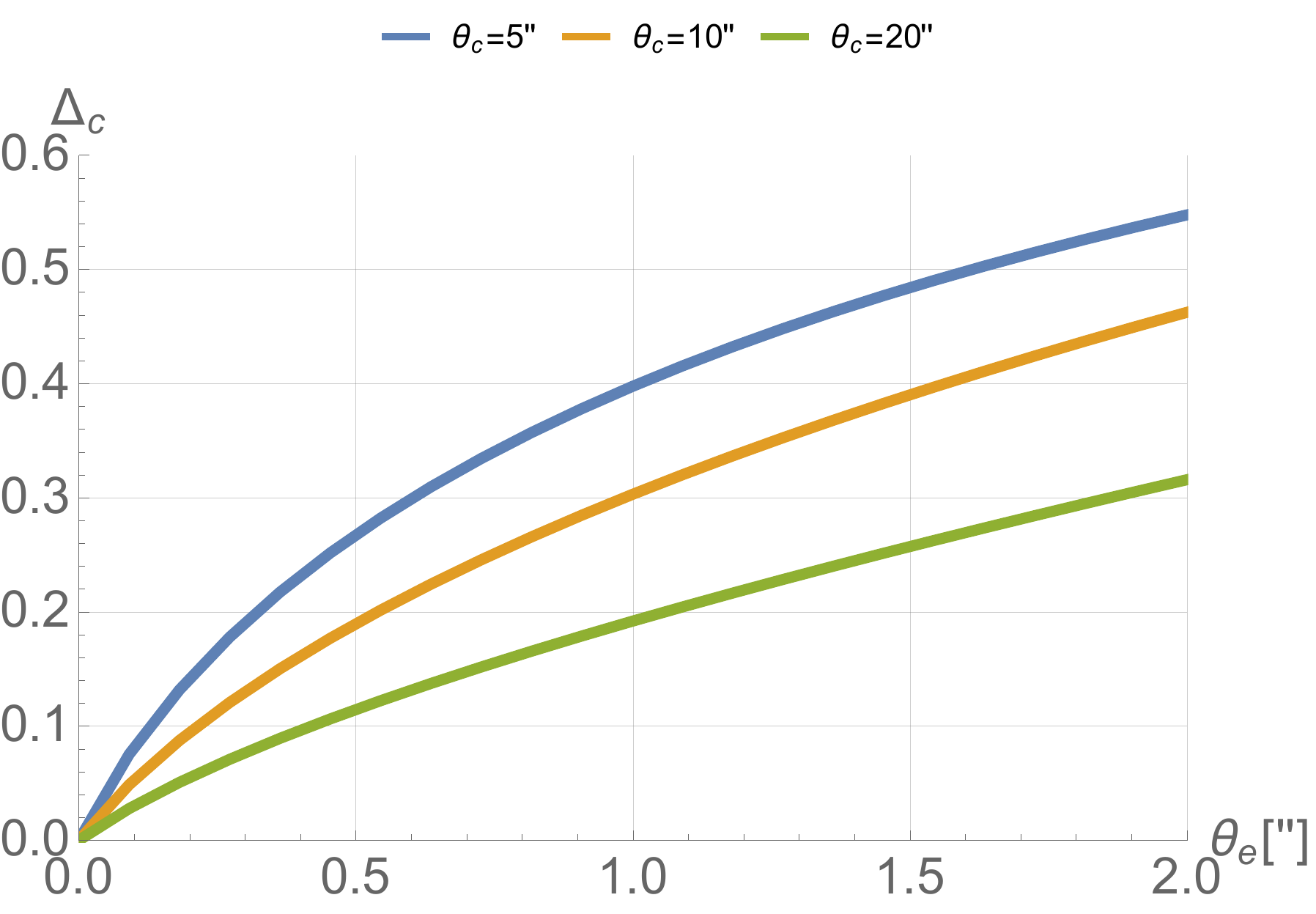}
 \caption{The finite-core correction $\Delta_{\rm c}$, modifying the MSD limit in the kinematics analysis (see Eqs.~(\ref{eq:Dc1}-\ref{eq:Dc2}) and text). 
Here we neglect the seeing, the aperture is defined to be a circular region of radius $1''$, and the null model has $\theta_{\rm E}=1''$.
 }
 \label{fig:kinDc}
\end{figure}

In Fig.~\ref{fig:kinDc} we plot $\Delta_{\rm c}$ vs. $\theta_{\rm e}$ for different values of $\theta_{\rm c}$. The null model is defined with $\theta_{\rm E}=1''$. The aperture is defined to be a circular region of radius $1''$ (a simplification of the aperture in~\cite{Birrer:2020tax}). For simplicity we neglect the seeing, setting the FWHM of $\mathcal{P}$ to zero. 

TDCOSMO systems typically have $\theta_{\rm E}\sim1''$, and from the imaging analysis we know that $\theta_{\rm c}\gtrsim5\theta_{\rm E}$ or so. 
%For the mock test of~\cite{Birrer:2020tax}, with $\theta_{\rm c}\approx20''$ and $\theta_{\rm e}=0.2''$, we can estimate $\Delta_{\rm c}\approx0.05$ and the  MSD limit is quite accurate. However, s
Some TDCOSMO systems have $\theta_{\rm e}\sim\theta_{\rm E}\sim1''$ (Fig.~16 in~\cite{Birrer:2020tax}); for such systems, $\Delta_{\rm c}$ can exceed 30\%. 
SLACS systems have even larger values of $\theta_{\rm e}$, some reaching $\theta_{\rm e}\sim2.5\theta_{\rm E}$, and Fig.~\ref{fig:kinDc} shows that the MSD limit may be violated at the $\mathcal{O}(1)$ level. The effect should be even more important for SLACS systems with resolved kinematics (see Figs.~15~\cite{Birrer:2020tax}). This is manifest, to some extent, in Fig.~B3 in~\cite{Birrer:2020tax}. In App.~\ref{a:msdkin} we estimate $\Delta_{\rm c}$ in more detail for resolved SLACS systems. 

The calculation in Fig.~\ref{fig:kinDc} does not include the effect of a finite PSF, velocity anisotropy, lens ellipticity, etc. In App.~\ref{a:msdkin} we repeat a similar calculation using a full mock system that includes all of these effects. The result of a full computation is compatible with that in Fig.~\ref{fig:kinDc} numerically to 50\% or so.

If a real physical core component is behind the lensing $H_0$ tension, then the kinematics constraints must be considered with care, because the MSD limit could introduce large systematic errors. In general, the breaking of the MSD manifests in a {\it smaller deviation} of $\sigma^{\rm P}$ from the null model: instead of $\delta\sigma^{\rm P}/\sigma^{\rm P}=-0.5\kappa_{\rm c}$ we have $\delta\sigma^{\rm P}/\sigma^{\rm P}=-0.5\kappa_{\rm c}(1-\Delta_{\rm c})$, with $\Delta_{\rm c}>0$. This calls into question the kinematics analysis of some TDCOSMO systems and certainly of resolved SLACS systems in~\cite{Birrer:2020tax}.

Finally, while we think that the kinematics data needs to be reconsidered, this is unlikely to change the conclusion that a core-MSD solution for the lensing $H_0$ tension is consistent with the data. Even if we conservatively take the MSD limit, Tab.~\ref{Tab:lenses} shows that the TDCOSMO systems driving the tension satisfy $2|\delta\sigma^{\rm P}/\sigma^{\rm P}|\gtrsim\delta H_0/H_0$ for all but PG1115, and there the inequality holds to 0.5$\sigma$ or so.

\section{Theoretical perspective}\label{s:th}
To explain the lensing $H_0$ tension, the ULDM soliton mass in the lens galaxy must be large enough. How much ULDM is needed, and how does this requirement compare to the soliton predicted by numerical and analytic considerations?%~\cite{Bar:2018acw}? 

Numerical simulations have shown that the soliton grows by accreting ULDM from the surrounding halo via gravitational dynamical relaxation, with a characteristic timescale 
\be\label{eq:taug}
\tau_\mathrm{g}&\approx&\frac{\sqrt{2}b}{12\pi^3}\frac{m^3\sigma^6}{G^2\rho_\chi^2\Lambda}  .
\ee
Here, $\rho_\chi$ is the density of ULDM, $\sigma$ is the velocity dispersion, $\Lambda$ is the Coulomb logarithm, and the numerical factor $b\approx0.7$ was calibrated in numerical simulations~\cite{Levkov:2018kau} (for recent analyses, see also~\cite{Eggemeier:2019jsu,Chen:2020cef,Schwabe:2020eac}). Below we will set $\sqrt{2}b\approx1$. 

A first estimate of the maximal mass of an ULDM soliton that could form in a galaxy can be obtained by calculating the ULDM mass contained inside the galactocentric radius $R_\mathrm{g}$ within which $\tau_\mathrm{g}(R_\mathrm{g})<t_{\rm gal}$, where $t_{\rm gal}$ is the age of the galaxy. Near this radial boundary we expect that $\rho_\chi\approx\alpha_\chi\rho=\alpha_\chi\rho_0/(1-\alpha_\chi)$, where $\rho$ is the total DM density (ULDM+non-ULDM) and $\rho_0$ is the background density in non-ULDM DM [$\alpha_\chi$ is the cosmological ULDM fraction defined in Eq.~(\ref{eq:alphax})]. We can thus estimate $R_\mathrm{g}$ from solving
\be\label{eq:Rg}
t_{\rm gal}&\approx&\frac{1}{12\pi^3}\frac{m^3\sigma^6(R_\mathrm{g})}{G^2\Lambda\alpha_\chi^2\rho^2(R_\mathrm{g})}  .
\ee
A rough upper bound on the mass of a soliton is then
\be\label{eq:Mmax} M&<&\alpha_\chi M_{\rm halo}(R_\mathrm{g})  .\ee

For an isothermal power-law halo with $\rho\propto R^{-2}$ and constant $\sigma\approx\sigma_0$, we have $\sigma^2_0\approx c_0GM(R_0)/R_0\approx c_0\,4\pi G\rho(R_0)R_0^2$ where we expect\footnote{See~\cite{2008gady.book.....B}, Ch.4.3. We keep track of the constant $c_0$ here because in a realistic scenario it could vary by $\mathcal{O}(1)$, contributing to the uncertainty in the relaxation estimate.} $c_0\approx1/2$. With this we have
\be R_\mathrm{g}^4&\approx&12\pi^3\frac{G^2\Lambda\alpha_\chi^2\rho^2(R_0)R_0^4}{m^3\sigma_0^6}t_{\rm gal}\\
&\approx&\frac{3\pi}{4}\frac{\Lambda\alpha_\chi^2}{c_0^2m^3\sigma_0^2}t_{\rm gal}  .\no
\ee
On the other hand, $M(R_\mathrm{g})\approx\sigma_0^2R_\mathrm{g}/(c_0G)$, so using Eq.~(\ref{eq:Mmax}) the soliton upper bound reads
\be\label{eq:Mmaxdat} M&<&\left(\frac{\alpha_\chi^2}{m}\right)^{3/4} \frac{1}{G}\qty(\frac{\sigma_0}{c_0})^{3/2}\left(\frac{3\pi}{4}\Lambda t_{\rm gal}\right)^{1/4}  .\ee

In Fig.~\ref{fig:sys} we show how the estimate of Eq.~(\ref{eq:Mmaxdat}) compare with the imaging and $H_0$ constraints. The upper bound is shown by the green, pink, and purple bands, corresponding to $\alpha_\chi=0.2,\,0.1$, and $0.05$, respectively. The upper and lower limits of each of the bands are obtained by setting $\sigma_0=\sigma^\mathrm{P}$ in Eq.~(\ref{eq:Mmaxdat}) and using the upper and lower uncertainty estimates for $\sigma^\mathrm{P}$ from Tab.~\ref{Tab:lenses}. The age of each lens galaxy [$t_{\rm gal}$ in Eq.~(\ref{eq:Mmaxdat})] is taken as the FRW time between $z=20$ and the lens redshift $z_\mathrm{l}$.

We truncate each constant-$\alpha_{\chi}$ band at small $m$ according to the cosmological constraints from Ref.~\cite{2021arXiv210407802L}. We also adhere, roughly, to the limit of~\cite{Kobayashi:2017jcf} by restricting to $\alpha_\chi\leq0.2$. Inspecting the result, it is clear that the cosmological constraints on $\alpha_\chi$ play an important role in the scenario. While the imaging constraints eliminate $m\gtrsim10^{-24}$~eV or so, the combination of the dynamical relaxation consideration with the cosmological bounds~\cite{Kobayashi:2017jcf,2021arXiv210407802L} disfavours $m\lesssim10^{-25}$. This defines the interesting parameter space of the model to a rather narrow window.

Apart from the dynamical relaxation upper bound, another consideration comes from the saturation of the growth of the soliton: while Eq.~(\ref{eq:Mmaxdat}) estimates the maximal amount of ULDM mass that is available for condensation into a soliton, it is possible that only a fraction of this total available mass would actually condense.
The soliton growth slows from $M\propto \left(t/\tau_\mathrm{g}\right)^{1/2}$ to $M\propto \left(t/\tau_\mathrm{g}\right)^{1/8}$ when the specific kinetic energy of the soliton (kinetic energy per unit mass, $(K/M)$) becomes comparable to the specific kinetic energy in the surrounding halo. Both the $M\propto t^{1/2}$ growth phase and its saturation into $M\propto t^{1/8}$ were observed in numerical simulations~\cite{Eggemeier:2019jsu,Chen:2020cef,Schwabe:2020eac}, and are consistent with the soliton--host halo relation originally discovered in~\cite{Schive:2014dra,Schive:2014hza}, and then shown to be equivalent to $(K/M)$ equilibration in~\cite{Bar:2018acw,Bar:2019bqz}. The reason for this saturation is that once the $(K/M)$ threshold is crossed, the velocity dispersion at the outskirts of the soliton, and thus the dynamical time scale $\tau_\mathrm{g}$, becomes dominated by the gravitational potential of the soliton itself. This causes $\tau_\mathrm{g}$ to depend on $M$ with larger $M$ corresponding to larger $\tau_\mathrm{g}$, leading to self-regulation of the growth rate. 

With the parameterization of Eq.~(\ref{eq:abapp}) we can compute the soliton specific kinetic energy,
\be\label{eq:K2Mab}
\left(\frac{K}{M}\right)_{\lambda}&=&\lambda^2\frac{\int \dd{r}r^2\left(\partial_r\chi_1\right)^2}{2\int \dd{r}r^2\chi_1^2}\no\\
&\approx&\lambda^2\frac{3a^2b^2\Gamma\left(2b\right)\Gamma\left(2b-\frac{1}{2}\right)}{\Gamma\left(2b+2\right)\Gamma\left(2b-\frac{3}{2}\right)}  .
\ee
In the limit of low-mass soliton, where the background gravitational potential completely dominates the structure of $\chi$, $(K/M)_\lambda$ is independent of the parameter $\lambda$ because the ULDM profile simply reflects the wave function of an ULDM particle bound in the external potential. Indeed, using the PL external potential in this limit gives $(K/M)_\lambda\approx A/2$, consistent with the virial theorem\footnote{To be precise, the large-$A/\lambda^2$ limit of Eq.~(\ref{eq:K2Mab}) gives $(K/M)_\lambda\to 0.454A$. The small mismatch from 1/2 can be expected given that Eq.~(\ref{eq:abapp}) is merely an approximation for the soliton.}. 
We can estimate the self-regulation threshold by letting the soliton mass grow until $(K/M)_\lambda$ starts to exceed the background-dominated result.

\begin{figure}
\centering
%\hspace*{-0.8cm}
 \includegraphics[scale=0.425]{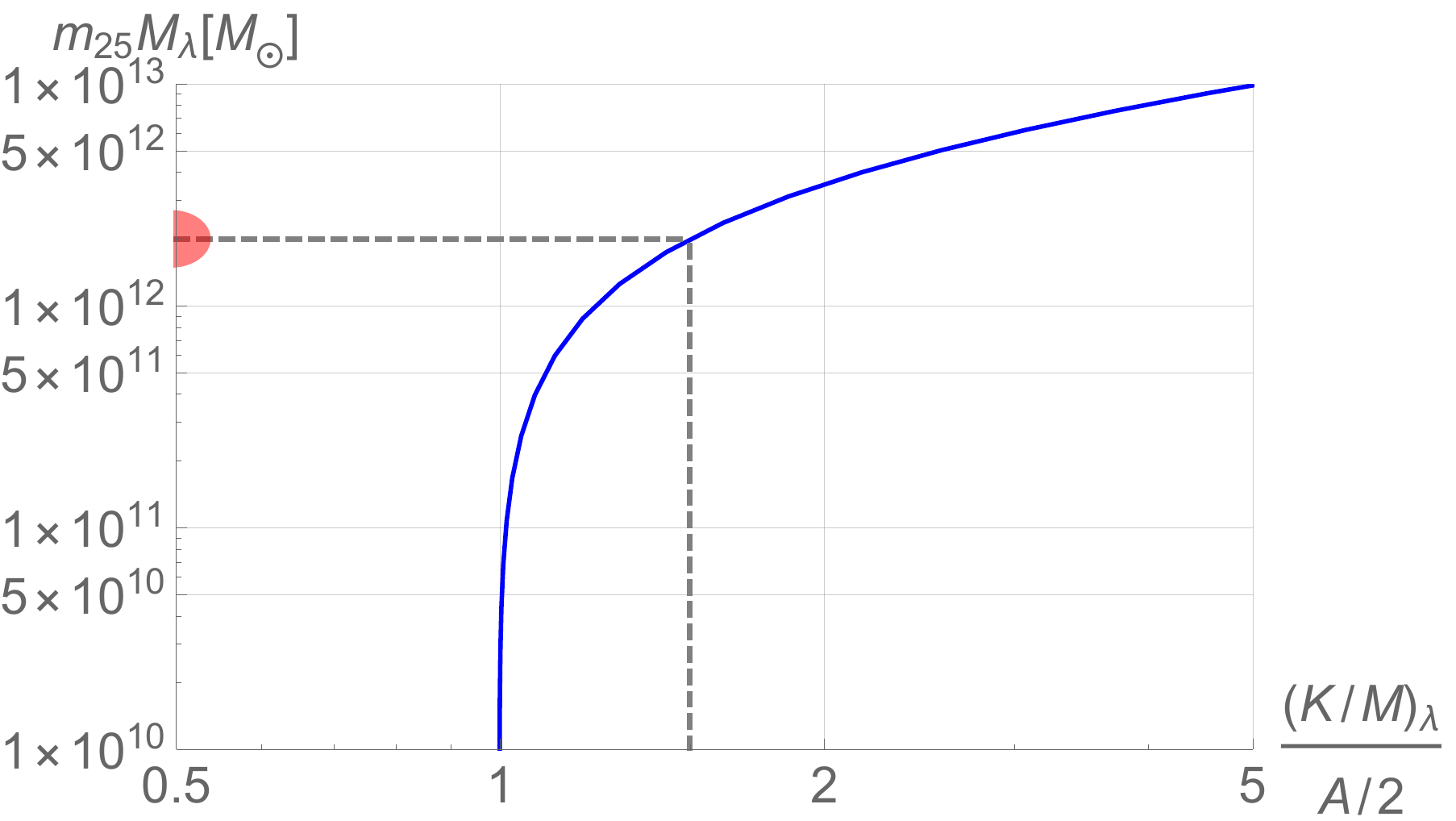} 
  \caption{Illustration of the soliton growth threshold, discussed in Sec.~\ref{s:th} [see text below Eq.~(\ref{eq:K2Mab})].}
 \label{fig:m25Mmax}
\end{figure}

In Fig.~\ref{fig:m25Mmax} we illustrate the growth saturation limit, computed for the system RXJ1131 from Tab.~\ref{Tab:lenses}. On the x-axis we plot $(K/M)_\lambda$ normalised to its asymptotic small-$M$ value. On the y-axis we plot the product $M_\lambda m_{25}$, where $m_{25}$ corresponds to the ULDM particle mass via $m_{25}=m/(10^{-25}~{\rm eV})$. As noted above, at small $M_\lambda$ the value of $(K/M)_\lambda$ becomes independent of $M_\lambda$ (or equivalently, of $\lambda$). As $M_\lambda$ increases, the soliton self-gravity begins to dominate $(K/M)_\lambda$. In Fig.~\ref{fig:m25Mmax}, we mark by a red dot the value of $M_\lambda$ at which $(K/M)_\lambda$ exceeds the small-$M$ result by 50\%. From Eq.~(\ref{eq:Mlam}), we know that in the self-gravitation limit the parameter $\lambda$ fixes the combination $M_\lambda m$; thus, the saturation limit also fixes the combination $M_\lambda m$. This is the reason why we use the product $M_\lambda m_{25}$ for the y-axis in Fig.~\ref{fig:m25Mmax}.

With some arbitrariness, we will estimate the growth saturation limit (roughly) by imposing, for each halo, $(K/M)_\lambda<1.5(A/2)$, similar to the illustration in Fig.~\ref{fig:m25Mmax}. The result of this calculation is shown by the thick dashed grey lines in Fig.~\ref{fig:sys}. 

For all of the systems of Tab.~\ref{Tab:lenses}, the growth saturation limit is weaker than or comparable to the dynamical relaxation time-scale constraint obtained with ULDM fraction $\alpha_\chi=0.2$. This suggests that for the range of $m$ plotted in Fig.~\ref{fig:sys}, ULDM solitons are still growing in the lens galaxies, and the limiting factor for the soliton mass may be the total ULDM mass available within the dynamically relaxed region of the halo.%, and not the self-regulated saturation of the soliton growth.

%%%%%%%%%%%%%%%%%%%
\section{Additional discussion}\label{s:add}
\subsection{Looking for a large-core soliton in near-by galaxies?}\label{ss:kinnear}
Stellar kinematics in well-resolved galaxies -- including, e.g., the Milky Way (MW) itself -- may provide additional constraints on ULDM. To our knowledge, the parametric region we consider here with $m\sim10^{-25}$~eV and ULDM fraction $\alpha_\chi\sim0.1$ has not been systematically studied yet. 

\begin{figure}
\centering
%\hspace*{-0.8cm}
 \includegraphics[scale=0.27]{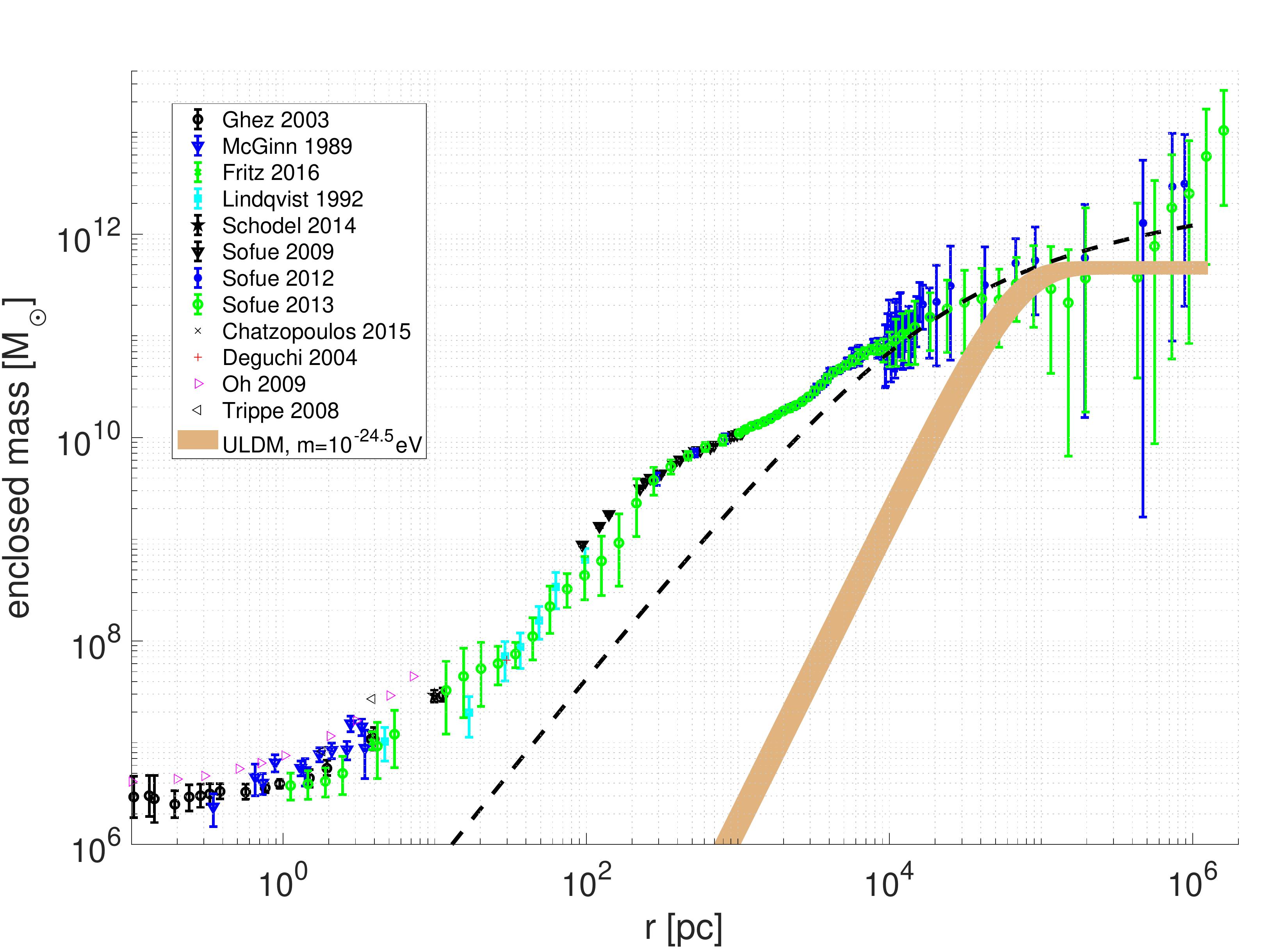}
 \caption{ULDM at $m=10^{-24.5}$~eV in the Milky Way? -- based on a collection of data referenced in~\cite{Bar:2018acw}.
 }
 \label{fig:solitonObs245}
\end{figure}

In a MW-like galaxy, the radius of the core would fall in the dozens of kpc range (comparable to the core radius for the massive elliptical lens galaxies in the cosmography analysis). 
Inwards of the core radius, ULDM would make a small perturbation to the total mass budget of the galaxy, and its presence may be difficult to detect. Near the core radius, however, ULDM may become observationally relevant. Fig.~\ref{fig:solitonObs245} illustrates how a soliton satisfying the soliton--halo relation~\cite{Schive:2014dra,Schive:2014hza} at $m=10^{-24.5}$~eV looks like in comparison to the observed kinematic mass budget of the MW. Clearly, a dedicated analysis of relevant data, notably from the GAIA mission~\cite{deSalas:2019pee,Prusti:2016bjo,Gaia:2021} could probe the scenario.

\subsection{Fluctuations and dynamical heating}
Ref.~\cite{Church:2018sro} estimated the dynamical heating due to ULDM fluctuations on MW disk stars (see also~\cite{Hui:2016ltb,Bar-Or:2018pxz}). For the case $\alpha_\chi=1$, where all of DM is ULDM, they quote a bound $ m \gtrsim \SI{e-22}{\electronvolt} $ by considering the vertical velocity dispersion of thick disk stars. Formally, in an infinite system, the rate of dynamical heating scales as $ m^{-3} \alpha_\chi^2 $, so a model with $\alpha_\chi\sim0.1$ and $m\sim10^{-25}$~eV could naively be thought to violate the bound. However, the MW is finite and in our model ULDM in the central few dozens of kpc (or even $\sim$100~kpc) is actually expected to be condensed in the coherent soliton (see Fig.~\ref{fig:solitonObs245}). In this region the stochastic heating analysis of~\cite{Hui:2016ltb,Bar-Or:2018pxz,Church:2018sro} is not valid. Instead of stochastic fluctuations, dynamical heating may still be transmitted to some extent to stars via core quasinormal mode fluctuations~\cite{Guzman2004,Veltmaat:2018dfz}. This analysis, for stellar orbits at the outskirts of the galactic disk, is yet to be done. (A related study~\cite{Marsh:2018zyw} considered soliton fluctuations heating a star cluster in a dwarf galaxy. These are very different regions in ULDM parameter space and system size.)

%%%%%%%%%%%%%%%%%%%
\section{Summary}\label{s:sum}
The possibility of a real tension between early- and late-type determinations of $H_0$ is exciting, and could signal a breakdown of $\Lambda$CDM~\cite{DiValentino:2021izs}. After all, the $\Lambda$CDM model is merely an effective theory. Gravitational lensing analyses, notably led by the TDCOSMO team, provide an important way to measure the local $H_0$. Accepting certain minimal assumptions about lens galaxy structure, the lensing analyses seem to reinforce the tension~\cite{Millon:2019slk}.

We follow up on the suggestion of Ref.~\cite{Blum:2020mgu}, that a core component in the density profile of lens galaxies would behave as an approximate internal mass sheet degeneracy (MSD) and could bring the lensing $H_0$ measurement down to the CMB value, solving the lensing part of the $H_0$ tension. A preliminary test of this proposition on the data was reported in TDCOSMO-IV~\cite{Birrer:2020tax}, finding a possible positive hint in the data. However, while~\cite{Birrer:2020tax} took an important step towards mitigating possible systematics related to the core-MSD proposal, they did not address the question of the physical origin of a core component. 

We explored ultralight dark matter (ULDM) as a concrete, well-motivated model that could naturally produce the required cores. If ULDM exists, then it is known to produce cores (``solitons") in the centre of galaxies, due to gravitational dynamical relaxation. We studied the lensing imprint of these cores and demonstrated that they could indeed address the lensing $H_0$ tension, if the ULDM particle mass is in the ballpark of $m\sim10^{-25}$~eV. Cosmological constraints~\cite{Kobayashi:2017jcf,Hlozek:2017zzf} imply that such light ULDM can only comprise $\lesssim$20\% of the total dark matter. This puts pressure on our scenario, because it limits the rate at which dynamical relaxation can operate and form the solitons. However, for ULDM abundance near this limit, the predicted cores are very close to the level required for $H_0$: clarifying this issue further would require numerical simulations that account for the background halo potential (tools of this type are already operational~\cite{Schive:2014dra,Schive:2014hza,Schwabe:2016rze,Veltmaat:2016rxo,Mocz:2017wlg,Veltmaat:2018dfz,Levkov:2018kau,Eggemeier:2019jsu,Chen:2020cef,Schwabe:2020eac}, but have so far been used to explore different parametric regions of ULDM). 

From a theoretical perspective, the required ULDM abundance could be realised via simple vacuum misalignment for an axion-like particle with a decay constant around the grand unification or string scale. 

Our study shows that strong galaxy lensing, combined with other cosmological probes like the CMB, could be sensitive to the presence of a subdominant component of dark matter in the form of ultralight fields or axions. It would be exciting if the lensing $H_0$ tension is the first hint for such fields, which could be the harbingers of otherwise inaccessible aspects of the UV theory. A promising path to test this idea is by dedicated kinematics studies, considering both massive elliptical galaxies of the type dominating the lensing analyses as well as near-by systems, including our own Milky Way.

%%%%%%%%%%%%%%%%%%%
\acknowledgments
We are grateful to N. Bar, E. Castorina, M. Simonovi\'c$^*$, and S. Suyu for comments on the manuscript, and to S. Birrer for useful comments and discussions and for guidance in using~\texttt{lenstronomy}. 
This work made use of the following public software packages: lenstronomy \cite{Birrer:2018,Birrer:2015}, emcee \cite{Foreman_Mackey:2013}, corner \cite{Foreman-Mackey2016}, astropy \cite{Robitaille:2013mpa,Price-Whelan:2018hus}, and FASTELL \cite{Barkana:1999}.
KB was supported by grant 1784/20 from the Israel Science Foundation, and is incumbent of the Dewey David Stone and Harry Levine career development chair. LT thanks R. Porto for hospitality at DESY Hamburg. The work was supported by the International Helmholtz-Weizmann Research School for Multimessenger Astronomy, largely funded through the Initiative and Networking Fund of the Helmholtz Association. \\

$^*$No hypergeometric functions were hurt in the preparation of this work.

\bibliography{ref}
\bibliographystyle{utphys}

\begin{appendix}
\section{Power-law background fitting formula}\label{a:PLan}
Here we consider how an external mass distribution affects the soliton profile. The background is taken to be a pure power-law (PL). Lensing analyses have often adopted this approximation, which leads to results for $H_0$ that are consistent with more realistic composite DM+stars halo models~\cite{Millon:2019slk}.
In a realistic analysis, the halo is axi-symmetric to accommodate quad geodesics, and we include axi-symmetry when we analyse mock data in App.~\ref{a:PLmock}. 
For simplicity, however, in modelling the impact of the external potential on the structure of the soliton we assume spherical symmetry. This approximation is justified by the disk galaxy study of Ref.~\cite{Bar:2019bqz}, which showed that the soliton remains nearly spherical even with significant a-sphericity of the background. 

The spherical PL density profile can be parameterised by
\be\label{eq:rhoPL}\rho_0(x)&=&\frac{\Sigma_\mathrm{c}}{D_\mathrm{l}\tilde\theta_\mathrm{E}}\frac{3-\gamma }{2\sqrt{\pi}}\frac{\Gamma\left(\frac{\gamma }{2}\right)}{\Gamma\left(\frac{\gamma -1}{2}\right)}\left(\frac{x}{D_\mathrm{l}\tilde\theta_\mathrm{E}}\right)^{-\gamma }  .\ee
This profile has two parameters: the PL slope $\gamma $ and the normalization, fixed here by  $\tilde\theta_\mathrm{E}$. (For a lensing model containing the PL $\rho_0$ alone, the parameter $\tilde\theta_{\rm E}$ would match the observable Einstein angle $\theta_{\rm E}$. This is no longer true once we consider composite models as in Eq.~(\ref{eq:msdc}).) The values of $D_\mathrm{l}$ and $\Sigma_\mathrm{c}$ are fixed by the system redshift and cosmology.
To simplify matters further we set $\gamma =2$, close to the slopes inferred for the galaxies in Tab.~\ref{Tab:lenses}. 
The external potential entering Eq.~(\ref{eq:SPE1}) is then given by:
\be\label{eq:PhiPL2}\Phi_{\rm ext}(r)&=&A\ln r\;+\;{\rm C}  ,\\
\label{eq:PhiPL2A}A&=&2G\Sigma_\mathrm{c}D_\mathrm{l}\tilde\theta_\mathrm{E}  .\ee
Note that the factor $\Sigma_\mathrm{c}D_\mathrm{l}=D_\mathrm{s}/(4\pi GD_\mathrm{ls})$ is independent of $H_0$.  
To gain some physical intuition, note that if we define $M_{\rm PL}(1/m)$ as the mass included in the PL profile up to a distance equal to the ULDM Compton radius $1/m$, then $A=GM_{\rm PL}(1/m)m$. Conveniently, for $\gamma =2$, $M_{\rm PL}(1/m)m$ is independent on $m$. 
The constant $\rm C$ in Eq.~(\ref{eq:PhiPL2}) is unimportant.  

Because $\Phi_{\rm ext}$ breaks the scale invariance of the self-gravitating soliton, the coefficients $a$ and $b$ in the approximation of Eq.~(\ref{eq:abapp}) now depend on the combination $A/\lambda^2$. We find that Eq.~(\ref{eq:abapp}) still provides a good fit for any value of $A/\lambda^2$, with the fitting formula: 
\be a(z)&=&0.23\sqrt{1+7.5z\tanh\left(1.5z^{0.24}\right)} \ ,\\
b(z)&=&1.69+\frac{2.23}{(1+2.2z)^{2.47}} \ ,\ee
where $z=A/\lambda^2$. 

\section{Power-law background: mock analysis}\label{a:PLmock}
Here we use the gravitational lens model software package~\texttt{lenstronomy}~\href{https://github.com/sibirrer/lenstronomy}{\faGithub}~\cite{Birrer:2018} to study the core-MSD soliton model in mock data analysis. Our main purpose is to check how well the simple $\delta_{\rm E}$ imaging error criterion described in Sec.~\ref{s:nobar} (see Eqs.~\eqref{eq:dE} and~\eqref{eq:dElam}) captures the observational constraints on the model. In addition, the implementation of the soliton core module in~\texttt{lenstronomy} would be useful to test the model directly against data in forthcoming work.

\begin{figure*}
	\centering
	\includegraphics[scale=0.35]{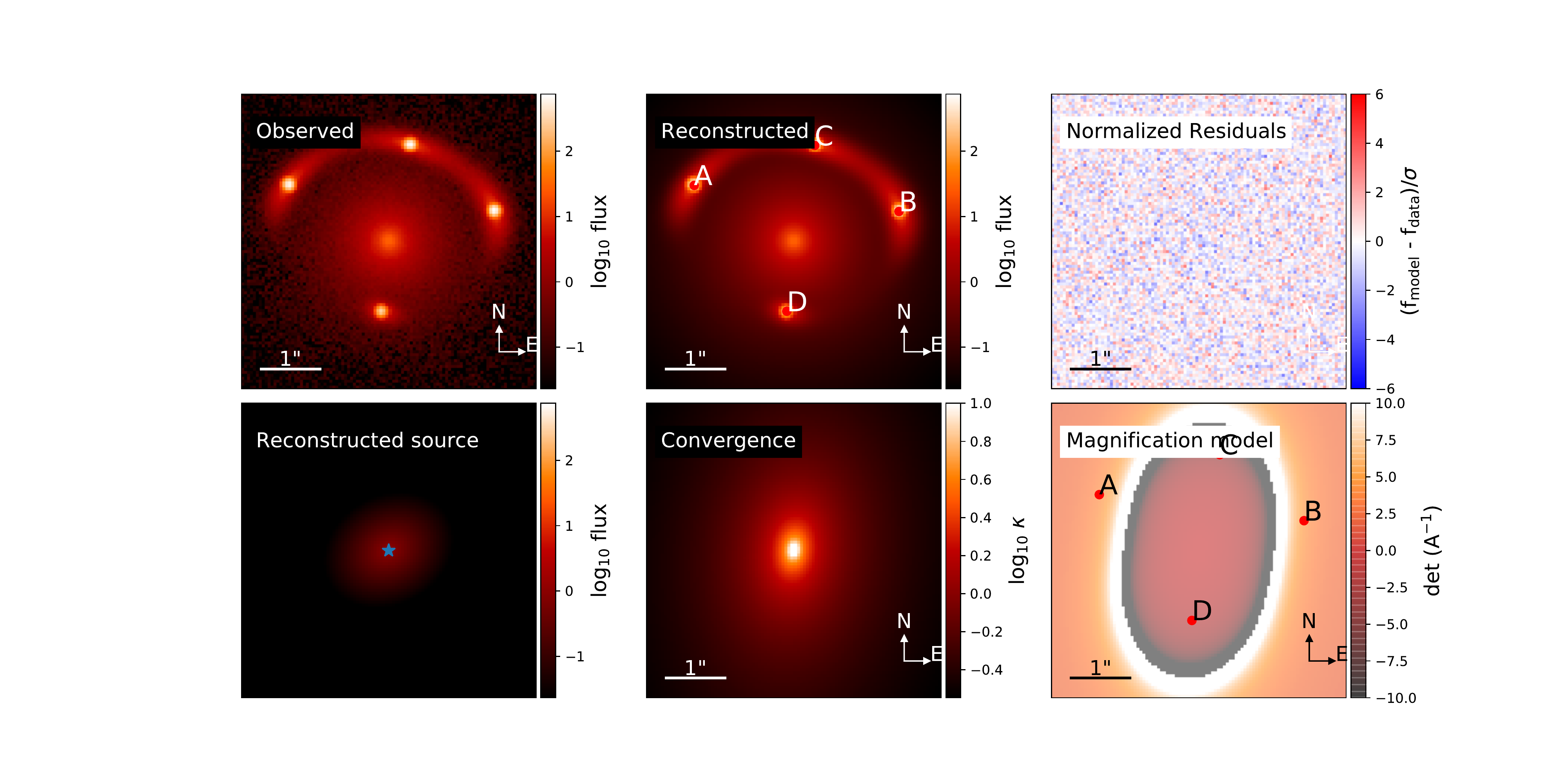}
	\caption{Mock image and reconstruction. The model used for inference is PL + ULDM core, with a gaussian prior of $H_0=67.4$~km/s/Mpc. Code:~\href{https://github.com/lucateo/ULDM-Strong-Lensing_H0/blob/main/Notebooks/Mock_analysis_uldm2uldm_H0_prior.ipynb}{\faGithub}.}
	\label{fig:mock_image}
\end{figure*}

The mock data is as follows. The truth model has the convergence of Eq.~\eqref{eq:msdc}, with $\kappa_{0}$ given by an elliptic PL profile (so as to produce a quad image) and $\kappa_{\rm c}=\kappa_\lambda$ of an ULDM soliton with $m=10^{-25}$~eV and $M=1.4 \times 10^{12}$~M$_\odot$. The parameters are chosen to produce an effective $\kappa_\lambda(\theta_\mathrm{E})\approx0.1$ and $ \theta_\mathrm{c} \approx 10'' $. The truth value of $H_0$ is set to $H_0=67.4$~km/s/Mpc, mimicking the CMB result~\cite{Akrami:2018vks}.  
In figure \ref{fig:mock_image} we show the mock alongside a reconstructed image, done by running the MCMC using the core-MSD model with a Gaussian prior on $H_0$ set at its truth value.

\begin{figure}[h!]
\centering
%\hspace*{-0.8cm}
 \includegraphics[scale=0.35]{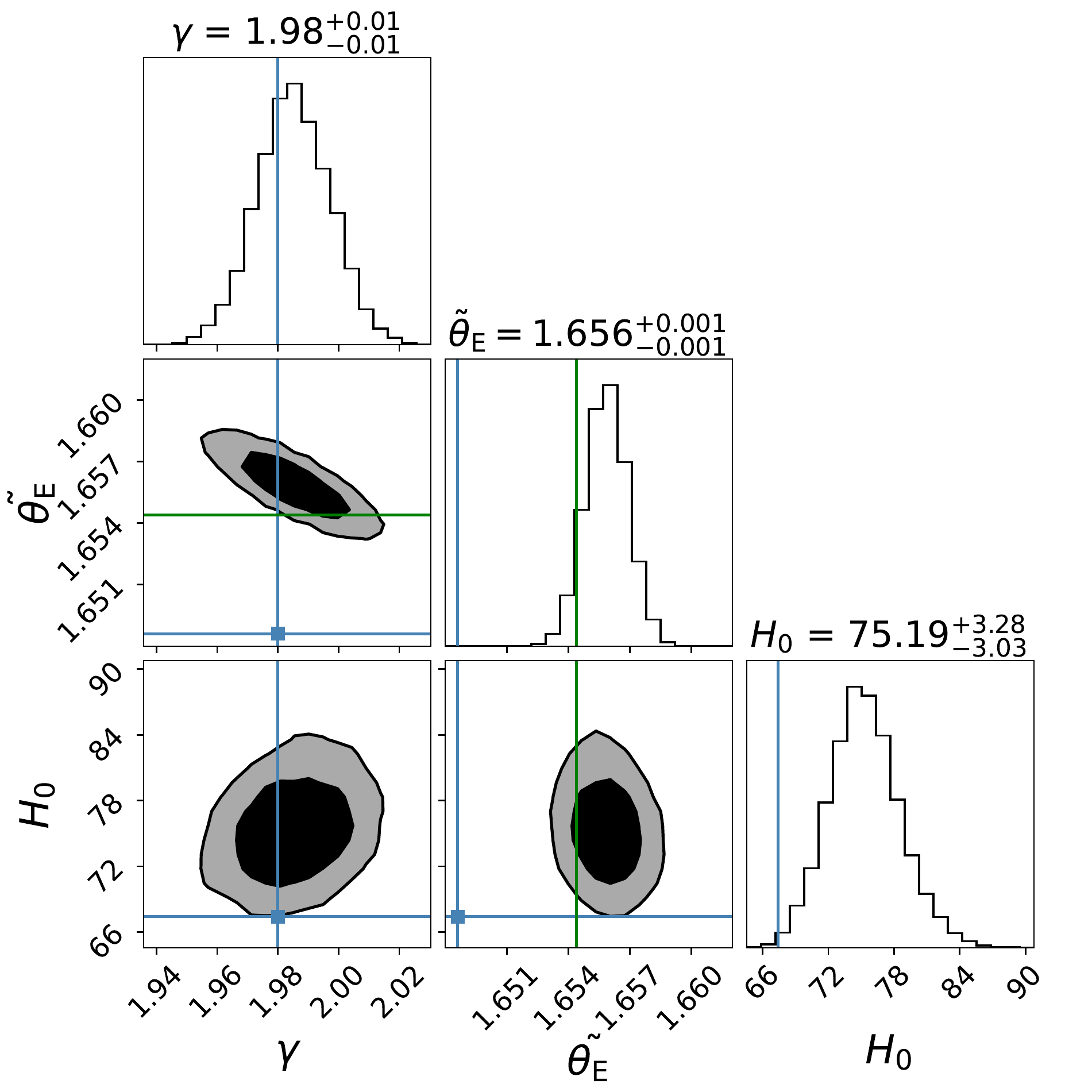}
 \caption{Lensing reconstruction and time delay analysis using mock data. Blue lines mark the truth values. The truth model is PL background + ULDM core, with $\kappa_\lambda(\theta_\mathrm{E}) \approx 0.1$. The truth value of $H_0$ used to produce the time delays is $H_{0,\rm true}=67.4$~km/s/Mpc. The model used in the inference is pure PL, without a core. The fit parameters are the PL slope $\gamma$, the Einstein angle $\tilde\theta_\mathrm{E}$, and the halo ellipticity $q$ (not shown in the plot). The PL fit converges on a false result $H_{0,\rm inferred}\approx75$~km/s/Mpc. Notice that this PL fit will try to converge to the true Einstein angle of the mock, which can be derived by solving $ \theta_{\rm E} = (1 - \kappa_\lambda(\tilde{\theta}_{\rm E}  )) \tilde{\theta}_{\rm E} + \alpha_\mathrm{c}(\theta_{\rm E}) $ where $ \tilde{\theta}_{\rm E} $ is the parameter we used to construct the mock; in green we show this $ \theta_{\rm E} $.
% Note that the PL $ \theta_\mathrm{E} $ parameter found by the MCMC is larger than the truth Einstein angle; this happens because, in the truth model, we parametrize the PL Einstein angle parameter as $ (1 - \kappa_{\lambda} (\theta_\mathrm{E}))\theta_\mathrm{E}  $; since we are dealing with an approximate MSD, this MCMC actually sees an Einstein angle who would be better parametrized in the true model by $ (1 - \kappa_\mathrm{eff}) \theta_\mathrm{E} $, where $ \kappa_\mathrm{eff} $ is an effective convergence which satisfies $ \kappa_{\lambda} (\theta_\mathrm{E}) < \kappa_\mathrm{eff} < \kappa_{\lambda}(0) $. To visually further clarify this point, we show in green the true value of $ \theta_\mathrm{E} $ had we decided to parametrize the Einstein angle as $ (1 - \kappa_{\lambda} (0))\theta_\mathrm{E} $ 
Code:~\href{https://github.com/lucateo/ULDM-Strong-Lensing_H0/blob/main/Notebooks/Mock_analysis_uldm2PL.ipynb}{\faGithub}.}
 \label{fig:mcmc1}
\end{figure}

To demonstrate the outcome of using an inference model which does not include a core component (the case of, e.g.~\cite{Suyu:2016qxx,Bonvin:2019xvn,Birrer:2018vtm,Chen:2019ejq,Wong:2019kwg,Millon:2019slk}), we run the MCMC using a pure (elliptic) PL. Fig.~\ref{fig:mcmc1} shows the posterior triangle plot obtained for this model. As expected, the MCMC converges to $H_0\approx75$~km/s/Mpc, in a good fit without detectable imaging residuals. A lensing analysis that does not utilise the core-MSD model would converge to this biased result.

\begin{figure}[h]
\centering
%\hspace*{-0.8cm}
 \includegraphics[scale=0.27]{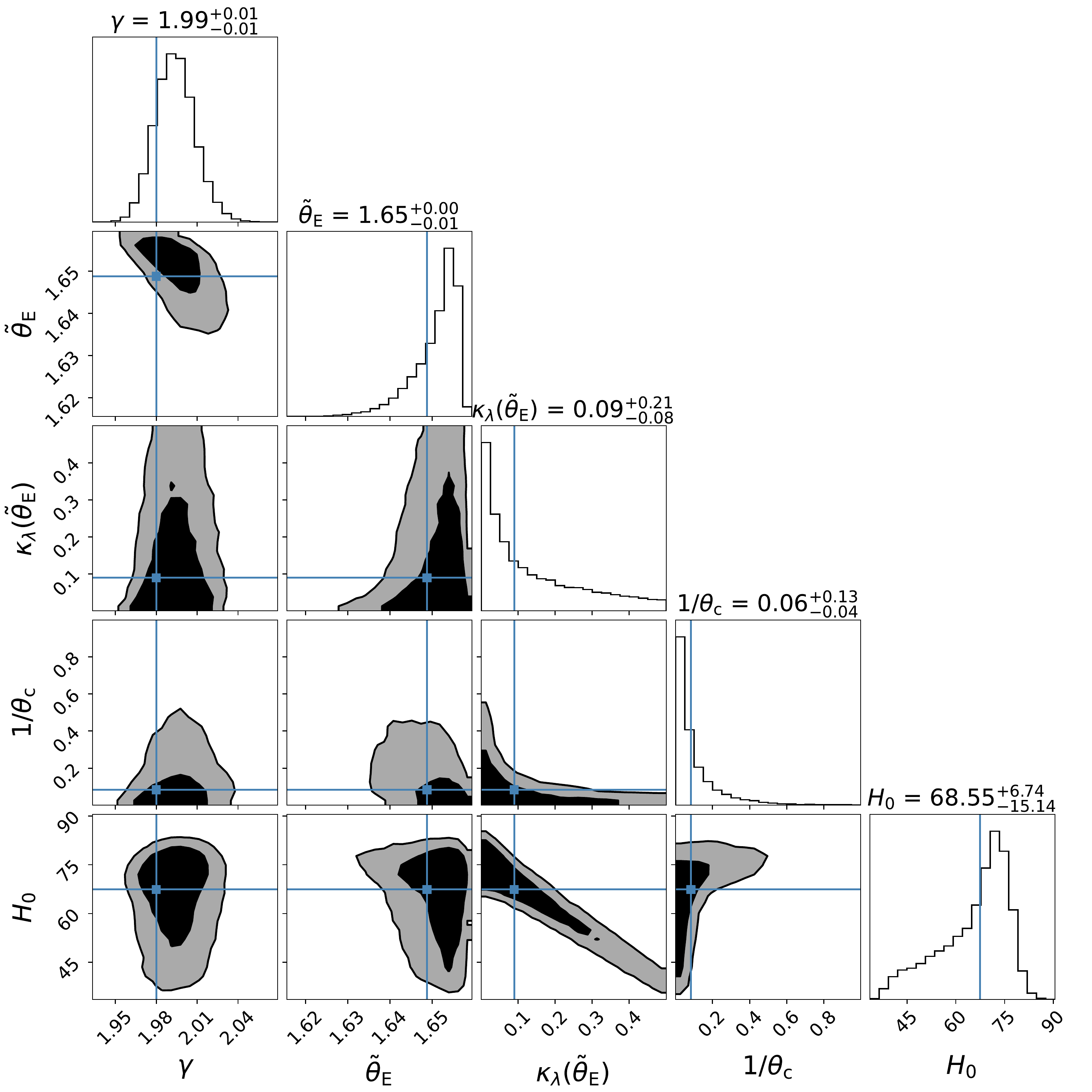}
  \includegraphics[scale=0.27]{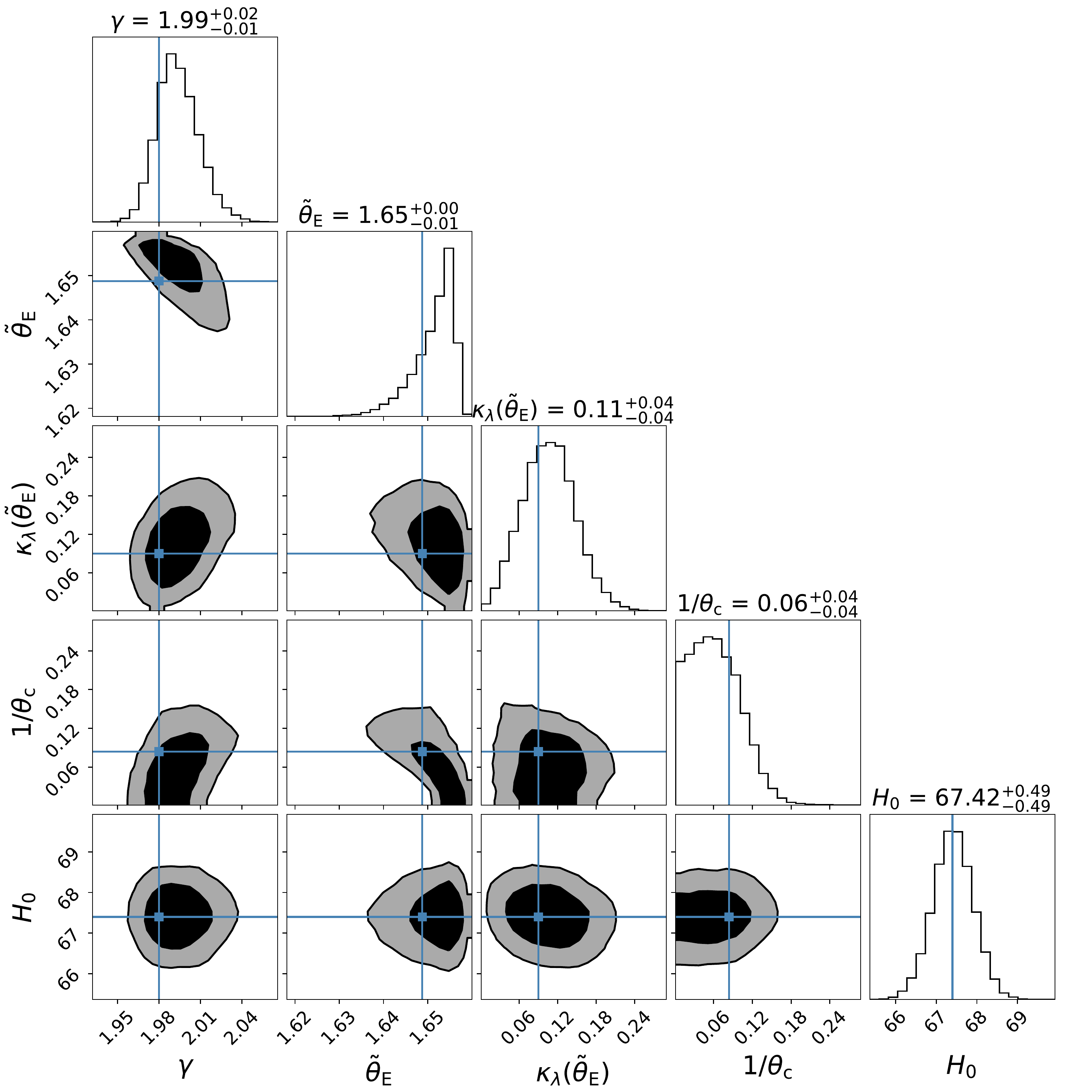}
 \caption{Lensing analysis for the same mock data as in Fig.~\ref{fig:mcmc1}, this time adding a core component to the fit. The blue lines correspond to the true values used for the mock.
 	{\bf Top:} Model inference with flat prior on $ H_0 $.  
 	We remark that the mock is consistent with a no-core solution; hence, the median value showed for $1/\theta_\mathrm{c}  $ is an artefact of the finite range of the prior.
 	The MSD is manifest by the broadening of the $H_0$ posterior distribution. Code:  \href{https://github.com/lucateo/ULDM-Strong-Lensing_H0/blob/main/Notebooks/Mock_analysis_uldm2uldm_No_H0_prior.ipynb}{\faGithub}. {\bf Bottom:} Same as in the top panel, this time adding a CMB prior on $H_0$. As expected, the no-core solution is now disfavoured. Code:~\href{https://github.com/lucateo/ULDM-Strong-Lensing_H0/blob/main/Notebooks/Mock_analysis_uldm2uldm_H0_prior.ipynb}{\faGithub}.}
 \label{fig:mcmc2}
\end{figure}

In the top panel of Fig.~\ref{fig:mcmc2} we re-run the MCMC, this time using the core-MSD model in the inference. (For convenience in the implementation, we use $1/\theta_{\rm c}$ and $\kappa_{\lambda}(\theta_{\rm E})$, rather than $m$ and $M$, as the sampling parameters in the fit.)
The MSD leads to a significant broadening of the $H_0$ posterior, corresponding to the $ \kappa_\lambda(\theta_\mathrm{E}) $-$ H_0 $ degeneracy. A low value of $H_0\approx60$~km/s/Mpc, accompanied by an $M\approx10^{12}$~M$_\odot$ soliton at $m\lesssim10^{-25}$~eV, produces a comparably good fit as the original $H_0\approx75$~km/s/Mpc model with a vanishing soliton (Fig.~\ref{fig:mcmc1}).

In the bottom panel of Fig.~\ref{fig:mcmc2} we repeat the exercise, this time adding an external CMB prior on $H_0$ coincident with the truth value of the mock. The posterior now converges to an upper limit of $ 1/\theta_\mathrm{c} \simeq 0.13 /(1'') $ at 95\% CL. This, together with the most probable value for  
$ \kappa_\lambda(\theta_\mathrm{E}) $, correspond to $ M \approx 9 \times 10^{10} $~M$_\odot$ and $ m \approx 2 \times 10^{-25}$ eV.

\begin{figure}[h]
	\centering
	\includegraphics[scale=0.45]{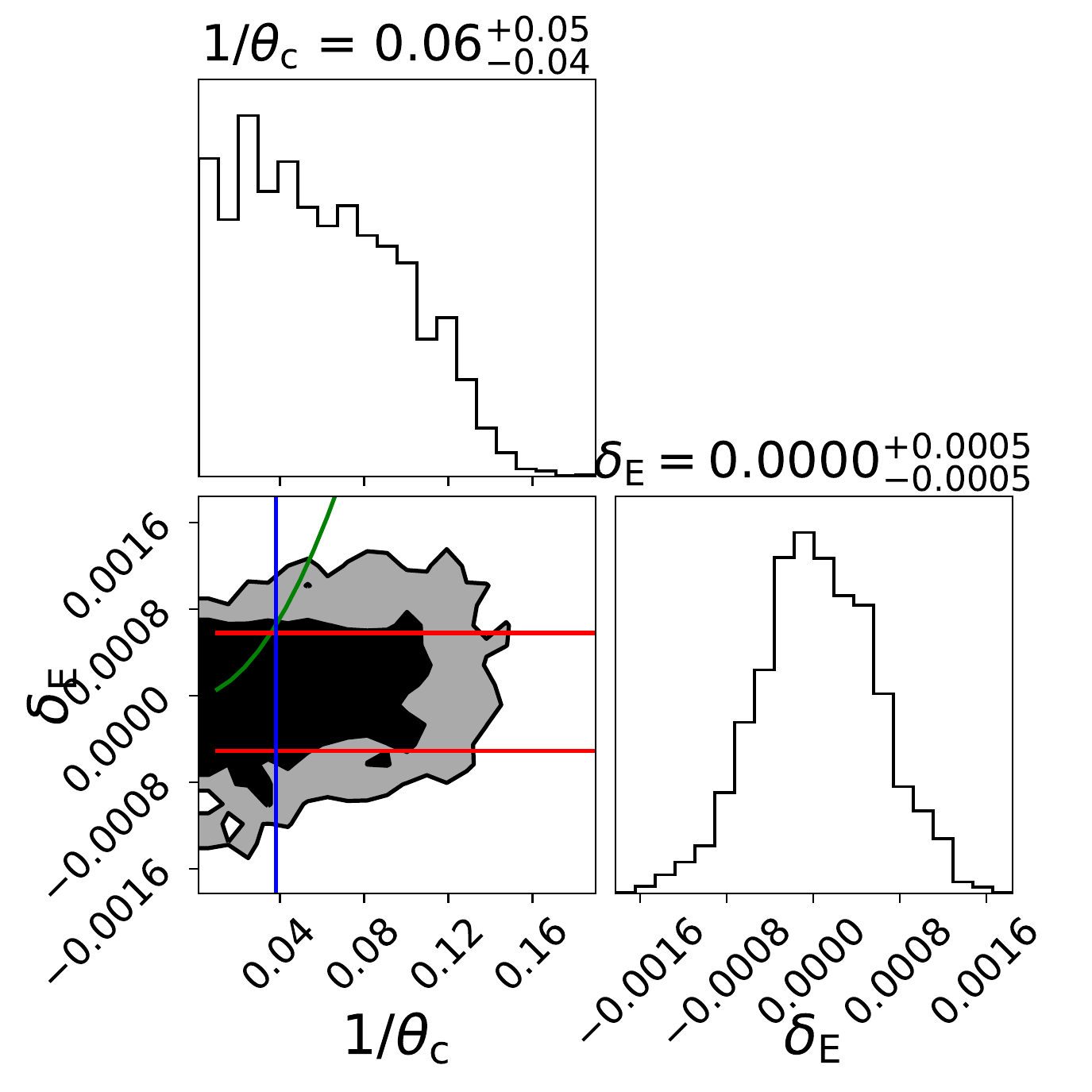}
	\caption{Triangle plot of $ \delta_\mathrm{E} $ and $ 1/\theta_\mathrm{c} $, calculated for a fixed $ \kappa_{\lambda}(\tilde{\theta}_{\rm E}) $. In green, the $ \delta_\mathrm{E} $ coming from Eq.~\eqref{eq:dElam}. Notice that the maximum $ \delta_\mathrm{E} $ allowed for our mock is $ |\delta_\mathrm{E}| \lesssim 0.0008 $ (horizontal red lines). Combining this with Eq.~\eqref{eq:dElam} would yield the naive bound  $ \theta_\mathrm{c} \gtrsim 25'' $ (blue vertical line, marking the intersection of the green and red lines). However, the region explored by the MCMC suggests that the more realistic bound is somewhat weaker, $ \theta_\mathrm{c} \gtrsim 10'' $. Code: \href{https://github.com/lucateo/ULDM-Strong-Lensing_H0/blob/main/Notebooks/Mock_analysis_uldm2uldm_H0_prior.ipynb}{\faGithub}.}
	\label{fig:deltaE_check}
\end{figure}

To study how well Eq.~\eqref{eq:dElam} approximates realistic imaging constraints on the soliton, in Fig.~\ref{fig:deltaE_check} we show $ \delta_\mathrm{E} $ as a function of $ 1/\theta_\mathrm{c} $, computed using Eq.~(\ref{eq:aE}) for a specific value of $ \kappa_{\lambda}(\theta_\mathrm{E}) $.
In this calculation, $ \alpha(\theta_\mathrm{E}) $ entering Eq.~(\ref{eq:aE}) is the deviation angle of the full core-MSD model, {\it computed at a fixed angle corresponding to the peak posterior value of $ \theta_\mathrm{E} $ found in the pure PL MCMC run} of Fig.~\ref{fig:mcmc1}. 
In green, we show the value given by Eq.~\eqref{eq:dElam}. We see that Eq.~\eqref{eq:dElam} leads to a  bound on $\theta_\mathrm{c} $ which is a factor of 2 or so stronger (that is, more conservative) than the MCMC bound. 

\section{MSD-breaking kinematics correction}\label{a:msdkin}
\begin{figure*}[]
	\centering
	\includegraphics[scale=0.45]{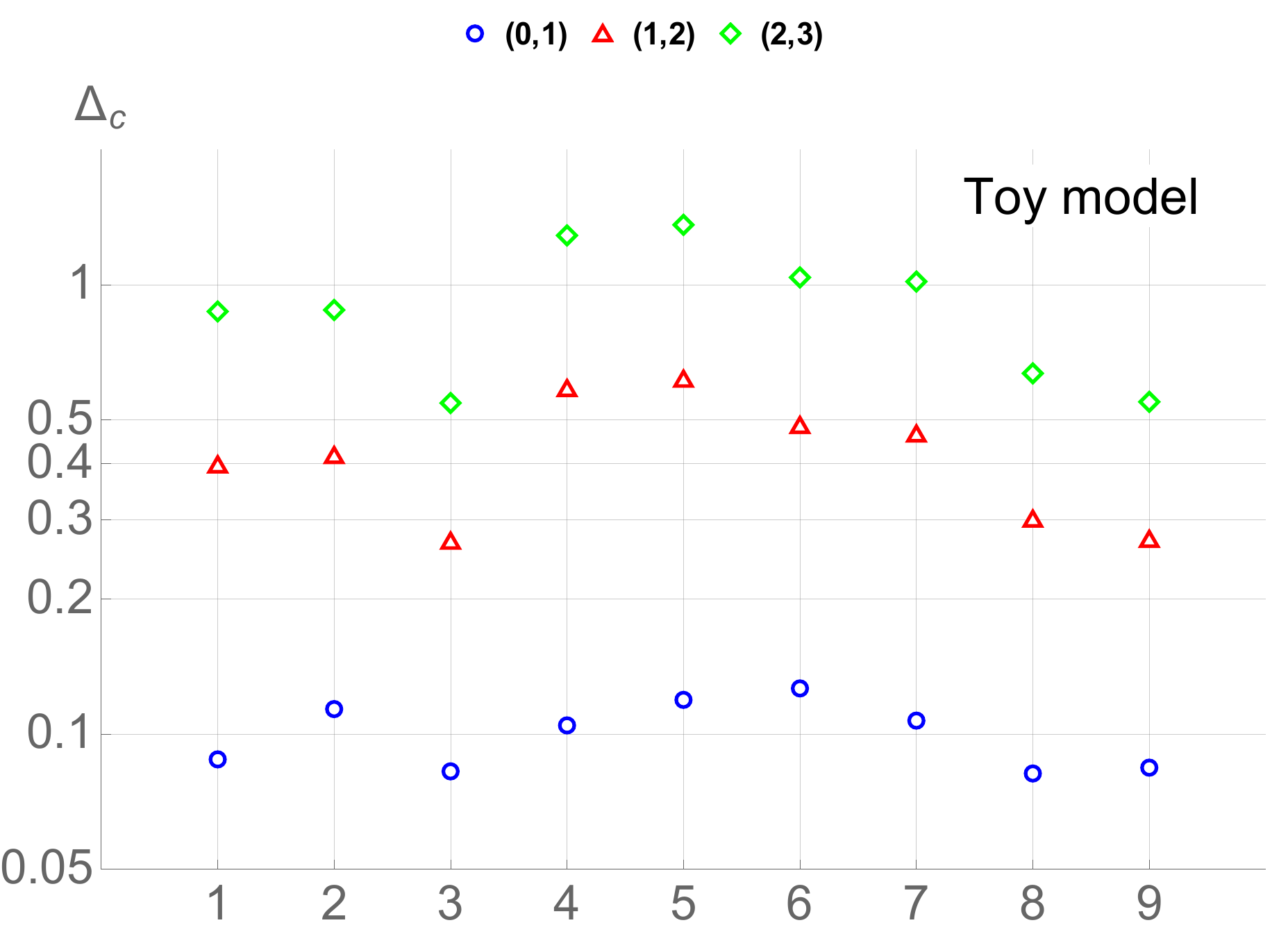}
	\includegraphics[scale=0.45]{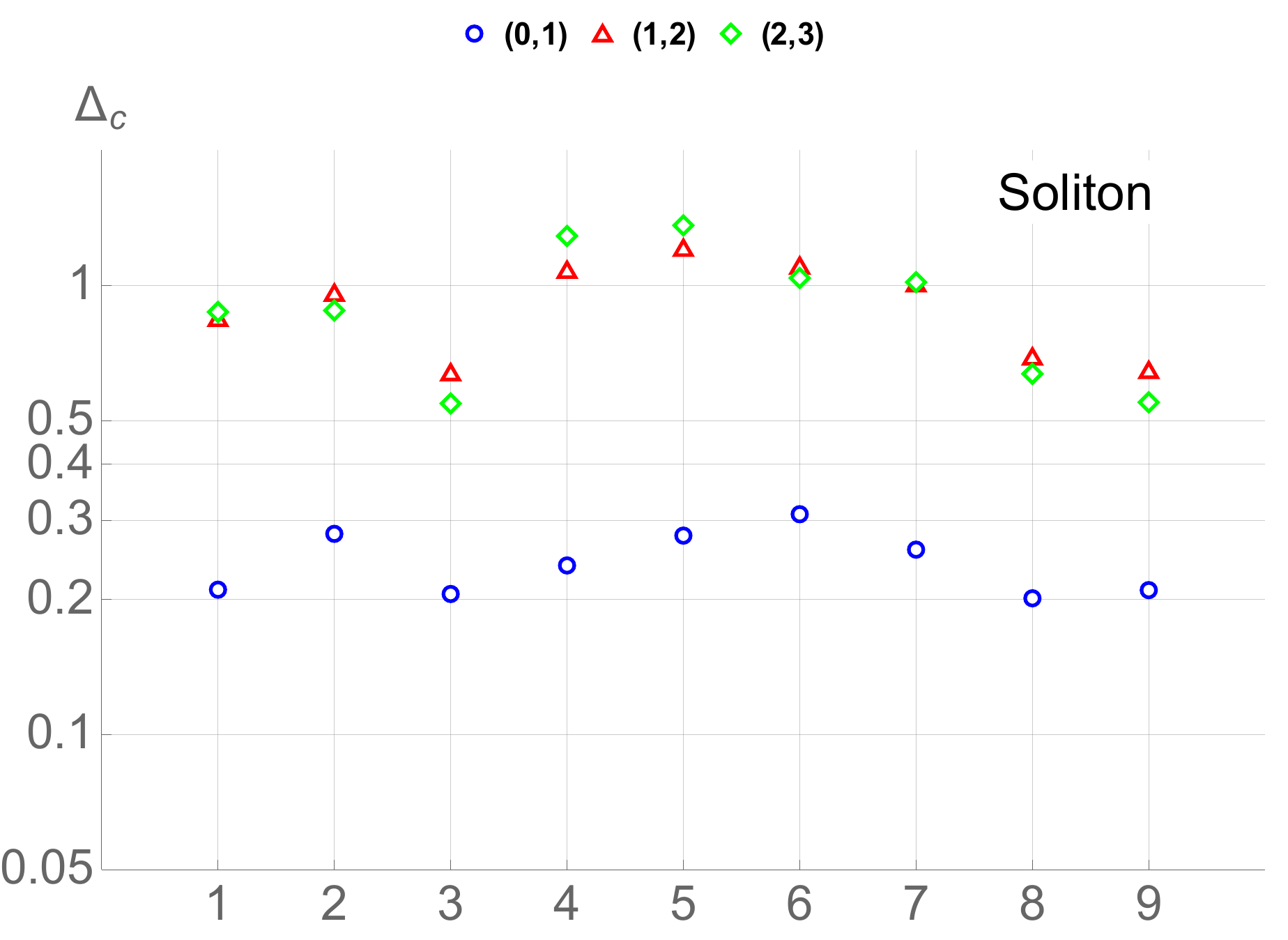}
	\caption{MSD-breaking kinematics correction $\Delta_{\rm c}$, computed semi-analytically (see Sec.~\ref{s:kin}) for model parameters mimicking the nine SLACS systems with resolved kinematics (see Fig.15 in~\cite{Birrer:2020tax}). Circle, triangle, and diamond markers correspond to the angular bins $(0'',1''),\,(1'',2''), (2'',3'')$, respectively. {\bf Left:} for the core toy model of~\cite{Birrer:2020tax}. {\bf Right:} for the ULDM soliton. Both models are defined with $\theta_{\rm c}=10\theta_{\rm e}$.}
	\label{fig:DcVIMOS}
\end{figure*}

In Fig.~\ref{fig:DcVIMOS} we show the MSD-breaking kinematics correction $\Delta_{\rm c}$, computed semi-analytically (see Sec.~\ref{s:kin}) for model parameters mimicking the nine SLACS systems of~\cite{Birrer:2020tax} with resolved kinematics data (see Fig.15 in~\cite{Birrer:2020tax}). The values of $\theta_{\rm e}$ and $\theta_{\rm E}$ for these system are taken from Tab.~E1 in~\cite{Birrer:2020tax}; from left to right, the systems in the plot are: SDSSJ1627-0053, SDSSJ2303+1422, SDSSJ1250+0523, SDSSJ1204+0358, SDSSJ0037-0942, SDSSJ0912+0029, SDSSJ2321-0939, SDSSJ0216-0813, SDSSJ1451-0239. Circle, triangle, and diamond markers correspond to the angular bins $(0'',1''),\,(1'',2''), (2'',3'')$, respectively. In the left panel we use the core toy model of~\cite{Birrer:2020tax}. In the right, we repeat the exercise for the physical ULDM soliton model. Both models are defined with $\theta_{\rm c}=10\theta_{\rm e}$.

In computing the effect for the core model of~\cite{Birrer:2020tax}, we use the fact that the density profile in this model matches (the square of) Eq.~(\ref{eq:abapp}). It follows that Eqs.~(\ref{eq:s2c}-\ref{eq:f2c}) are still valid for this model. The only adjustment needed is to set  $b=3/4$ for the toy model (compared to $b\approx3.9$ for a self-gravitating soliton). The parameter $\theta_{\rm c}$ has the same role in both cases. Fig.~\ref{fig:DcVIMOS} shows that for small apertures, the toy $\Delta_{\rm c}$ is roughly half that of a soliton defined at the same $\theta_{\rm c}$. 

\begin{figure}[h]
	\centering
	\includegraphics[scale=0.55]{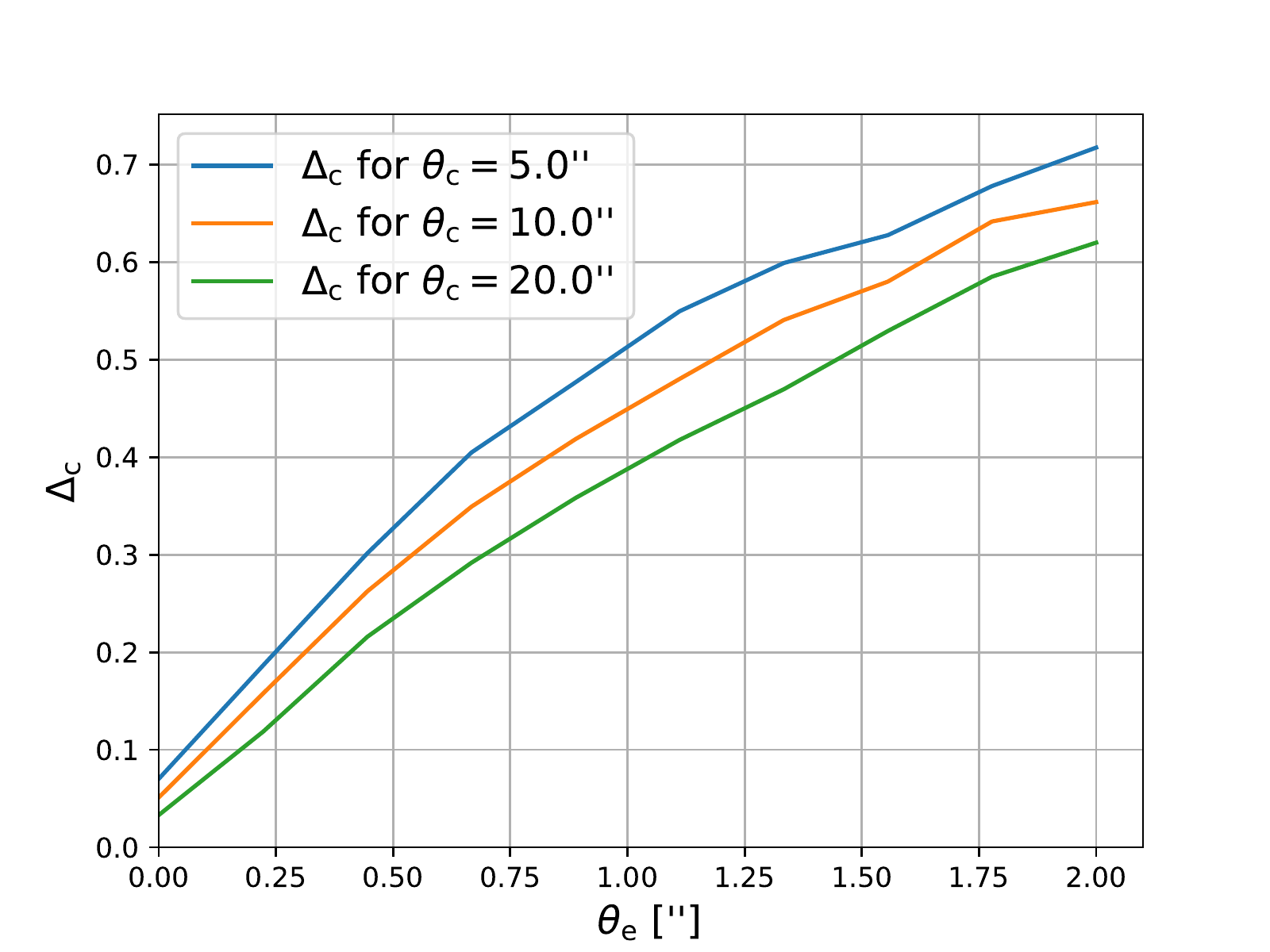}
	\caption{MSD-breaking kinematics correction $\Delta_{\rm c}$, computed numerically for the mock system including velocity anisotropy, lens ellipticity, PSF effects, and realistic aperture definitions~\href{https://github.com/lucateo/ULDM-Strong-Lensing_H0/blob/main/Notebooks/Velocity_dispersion.ipynb}{\faGithub}. Compare this result to the semi-analytic result in Fig.~\ref{fig:kinDc}.}
	\label{fig:Dccheck}
\end{figure}

We can calculate $\Delta_{\rm c}$ numerically, including effects like velocity anisotropy, lens ellipticity, PSF, and realistic aperture definitions that were lacking above and in Sec.~\ref{s:kin}. Fig.~\ref{fig:Dccheck} shows a full numerical computation of $\Delta_{\rm c}$, calculated directly from the definition Eq.~(\ref{eq:Dc1})~\href{https://github.com/lucateo/ULDM-Strong-Lensing_H0/blob/main/Notebooks/Velocity_dispersion.ipynb}{\faGithub}. The mock is defined with $\theta_{\rm E}=1.2''$, compared to $\theta_{\rm E}=1''$ in Fig.~\ref{fig:kinDc}. This means that if the PSF, aperture, anisotropy, and axi-symmetry effects were not important, we would expect $\Delta_{\rm c}$ computed from the mock in Fig.~\ref{fig:Dccheck} to be smaller by a factor $\approx0.83$ compared to Fig.~\ref{fig:kinDc}. In practice, with all of the above effects included, $\Delta_{\rm c}$ in Fig.~\ref{fig:Dccheck} is slightly larger. The parametric dependence on $\theta_{\rm e}$ and the rough size of the effect are well reproduced. Lastly, we verified that the full numerical procedure coincides very accurately (to $\mathcal{O}(1\%)$) with the analytical calculation when lens ellipticity and velocity anisotropy are set to zero.

\end{appendix}

\end{document}